\documentclass[10pt,conference,letterpaper]{IEEEtran}
\pdfoutput=1

\usepackage{mathpartir}

\usepackage{stmaryrd}

\usepackage{textcomp}                        
\usepackage[clock]{ifsym}

\usepackage{booktabs}
\usepackage{dcolumn}
\newcolumntype{.}{D{.}{.}{-1}}

\usepackage{nicefrac}

\def\clap#1{\hbox to 0pt{\hss#1\hss}}

\def\mathrlap{\mathpalette\mathrlapinternal}
\def\mathclap{\mathpalette\mathclapinternal}

\def\mathrlapinternal#1#2{%
\rlap{$\mathsurround=0pt#1{#2}$}}
\def\mathclapinternal#1#2{%
\clap{$\mathsurround=0pt#1{#2}$}}

\usepackage[flushleft,alwaysadjust]{paralist}

\usepackage{amsmath}
\usepackage{amssymb}

\usepackage{fancyvrb}
\usepackage{listings}

\DefineShortVerb{\|}
\SaveVerb{dollar}|$|\SaveVerb{dollar}|$|
\UndefineShortVerb{\|}  
\newcommand{\vardollar}{\UseVerb{dollar}}

\usepackage{dblfloatfix}

\usepackage[bottom,flushmargin]{footmisc}

\usepackage{subfigure}

\usepackage{url}

\usepackage{cmtt}

\usepackage{tikz}
\usetikzlibrary{arrows}
\usetikzlibrary{snakes}

\usepackage{fncylab}

\newcommand{\lbl}[3]{%
  \labelformat{#1}{#2}\refstepcounter{#1}\label{#3}
  \makeatother\csname p@#1\endcsname\makeatletter}

\usepackage{colortbl}

\newcolumntype{H}{>{\columncolor{black}\color{white}}c}
\newcommand{\colhd}[1]{\multicolumn{1}{H}{\col{#1}}}
\newenvironment{littbl}{%
  \renewcommand{\arraystretch}{0.7}
  \renewcommand{\arraycolsep}{1pt}}{}

\newcommand{\col}[1]{\mathsf{#1}}
\newcommand{\kind}[1]{\sql{#1}}

\newcommand{\Pathfinder}{\emph{Path\-finder}}

\newcommand{\env}{\ensuremath{\Gamma}}
\newcommand{\LOOP}{\ensuremath{\col{loop}}}
\newcommand{\map}{\ensuremath{\col{map}}}

\newcommand{\SELECT}{\sql{SELECT}}
\newcommand{\DISTINCT}{\sql{DISTINCT}}

\newcommand{\FROM}{\sql{FROM}}
\newcommand{\WHERE}{\sql{WHERE}}

\newcommand{\ORDERBY}{\sql{ORDER\,BY}}

\newcommand{\doc}{\ensuremath{\col{doc}}}
\newcommand{\select}{\ensuremath{\sigma}}
\newcommand{\cross}{\ensuremath{\times}}
\newcommand{\distinct}{\ensuremath{\delta}}
\newcommand{\num}{\ensuremath{{\scriptstyle\#}}}
\newcommand{\rank}{\ensuremath{\varrho}}
\newcommand{\attach}{{\textit{\scriptsize@}}}
\newcommand{\union}{\ensuremath{\cup}}

\DeclareMathOperator{\ser}{%
  \begin{tikzpicture}[baseline=0.5ex,scale=0.5,x=1mm,y=5mm,inner sep=1pt,thin]
    \draw[-o] (0,0) -- (0,1);
    \useasboundingbox (-2,0) rectangle (2,1);
  \end{tikzpicture}
}

\newcommand{\anticip}[1]{%
  \tikz[baseline=(X.base)] 
    \node[draw,rounded corners,dash pattern=on 2pt off 1pt,
          outer sep=1.5pt,inner sep=1.5pt] (X) {#1};}

\newcommand{\ndots}{\relax\ensuremath{\mathinner{\ldotp\ldotp}}}

\newcommand{\is}{\shortleftarrow}

\newcommand{\icols}{\ensuremath{\var{icols}}}
\newcommand{\const}{\ensuremath{\var{const}}}
\newcommand{\key}{\ensuremath{\var{key}}}
\newcommand{\set}{\ensuremath{\var{set}}}

\newcommand{\circled}[1]{%
  \text{\textcircled{\raisebox{-0.1ex}{\scriptsize#1}}}}
         
\newcommand{\pureXML}{\mbox{pure\XML\texttrademark{}}}

\newcommand{\XML}   {X\kern-0.07emM\kern-0.07emL}
\newcommand{\XPath} {X\kern-0.07emPath}
\newcommand{\XQuery}{X\kern-0.12emQue\-ry}
\newcommand{\SQL}   {S\kern-0.07emQ\kern-0.07emL}

\newcommand{\xquery}[1]{\texttt{#1}}
\newcommand{\sql}   [1]{\texttt{#1}}
\newcommand{\var}   [1]{\ensuremath{\mathit{#1}}}

\begin{document}

\title{%
 XQuery{} Join Graph Isolation\\[-0.5ex]
{\bfseries\large (Celebrating 30+ Years of \XQuery{} Processing Technology)$^1$}\vskip-5mm}

\author{%
  \begin{tabular*}{\textwidth}{@{\extracolsep\fill}ccccc}
                  &
    Torsten Grust &
    Manuel Mayr   &
    Jan Rittinger &
  \end{tabular*}
  \\[10pt]
  \fontsize{10}{10}\selectfont\rmfamily\itshape 
  Eberhard Karls Universit\"at T\"ubingen \\
  T\"ubingen, Germany 
  \\[6pt]
  \fontsize{9}{9}\selectfont\ttfamily\upshape 
  torsten.grust | manuel.mayr | jan.rittinger@uni-tuebingen.de
}

\maketitle
\thispagestyle{plain}

\let\ttfamily\mttfamily

\begin{abstract}

A purely relational account of the true \XQuery{} semantics can turn any
relational database system into an \XQuery{} processor. Compiling nested
expressions of the fully compositional \XQuery{} language, however, yields
odd algebraic plan shapes featuring scattered distributions of join
operators that currently overwhelm commercial \SQL{} query optimizers.

This work rewrites such plans before submission to the relational database
back-end.  Once cast into the shape of join graphs, we have found
off-the-shelf relational query optimizers---the B-tree indexing subsystem
and join tree planner, in particular---to cope and even be autonomously
capable of ``reinventing'' advanced processing strategies that have
originally been devised specifically for the \XQuery{} domain, \emph{e.g.},
\XPath{} step reordering, axis reversal, and path stitching. Performance
assessments provide evidence that relational query engines are among the
most versatile and efficient \XQuery{} processors readily available today.
\end{abstract}

\footnotetext[1]{This article is an extended version of a paper published in the Proceedings of the 25th IEEE International Conference on Data Engineering (ICDE 2009).}
\stepcounter{footnote}

\section{Introduction}

\SQL{} query optimizers strive to produce query plans whose primary
components are \emph{join graphs}---bundles of relations interconnected
by join predicates---while a secondary, peripheral \emph{plan tail}
performs further filtering, grouping, and sorting. Plans of this
particular type are subject to effective optimization strategies that,
taking into account the available indexes and applicable join methods,
derive equivalent join trees, ideally with a left-deep profile to enable
pipelining. For more than 30~years now, relational query processing
infrastructure has been tuned to excel at the evaluation of plans of
this shape.

\SQL{}'s rather rigid syntactical block structure facilitates its
compilation into join graphs. The compilation of \emph{truly
compositional} expression-oriented languages like \XQuery{}, however,
may yield plans of unfamiliar shape~\cite{XQueryOnSQLHosts}. The
arbitrary nesting of \xquery{for} loops (iteration over ordered item
sequences), in particular, leads to plans in which join and sort
operators as well as duplicate elimination occur throughout. Such plans
overwhelm current commercial \SQL{} query optimizers: the numerous
occurrences of sort operators block join operator movement, effectively
separate the plan into fragments, and ultimately lead to unacceptable
query performance.

Here, we propose a plan rewriting procedure that derives join graphs
from plans generated by the \XQuery{} compiler described 
in~\cite{XQueryOnSQLHosts}.  The \XQuery{} order and duplicate semantics 
are preserved.  The resulting plan may be equivalently expressed as a
single \SELECT-\DISTINCT-\FROM-\WHERE-\ORDERBY{} block to be submitted for
execution by an off-the-shelf RDBMS.  The database system then evaluates
this query over a schema-oblivious tabular encoding of \XML{} documents to
compute the encoding of the resulting \XML{} node sequence (which may then
be serialized to yield the expected \XML{} text).

In this work we restrict ourselves to the \XQuery{}~Core fragment, defined
by the grammar in Fig.~\ref{grammar}, that admits the orthogonal nesting
of \xquery{for} loops over \XML{} node sequences (of type \xquery{node()*}),
supports the 12 axes of \XQuery's \emph{full axis} feature, arbitrary
\XPath{} name and kind tests, as well as general comparisons in conditional
expressions whose \xquery{else} clause yields the empty sequence
\xquery{()}. As such, the fragment is considerably more expressive than the
widely considered \emph{twig} queries \cite{HolisticTwigJoin,Twig2Stack} and
can be characterized as \XQuery{}'s data-bound ``workhorse'': \XQuery{}
uses this fragment to collect, filter, and join nodes from
participating \XML{} documents.

\begin{figure}
\centering\small
$
\begin{array}{rclr}
    \var{Expr}
  & \to & \multicolumn{2}{l}{%
    \xquery{for}~\var{\vardollar\!VarName}~\xquery{in}~\var{Expr}~\xquery{return}~\var{Expr}} \\
  & |   & \multicolumn{2}{l}{%
    \var{\vardollar\!VarName}} \\
  & |   & \multicolumn{2}{l}{%
    \xquery{if}~\xquery(\var{BoolExpr}\xquery)~\xquery{then}~\var{Expr}~\xquery{else}~\xquery{()}} \\
  & |   & \multicolumn{2}{l}{%
    \xquery{doc(}\var{StringLiteral}\xquery)} \\
  & |   & \multicolumn{2}{l}{%
    \var{Expr}\,\xquery{/}\,\var{ForwardAxis}~\var{NodeTest}} \\
  & |   & \multicolumn{2}{l}{%
    \var{Expr}\,\xquery{/}\,\var{ReverseAxis}~\var{NodeTest}} \\
    \var{BoolExpr}
  & \to & \multicolumn{2}{l}{%
    \var{Expr}} \\
  & |   & \multicolumn{2}{l}{%
    \var{Expr}~\var{GeneralComp}~\var{Literal}} \\
    \var{GeneralComp}
  & \to & \xquery{=}\,|\,\xquery{!=}\,|\,\xquery{<}\,|\,\xquery{<=}\,|\,\xquery{>}\,|\,\xquery{>=} & [60] \\
    \var{ForwardAxis}
  & \to & \xquery{descendant::}\,|\,\xquery{following::}\,|\,\cdots & [73] \\
    \var{ReverseAxis}
  & \to & \xquery{parent::}\,|\,\xquery{ancestor::}\,|\,\cdots & [76] \\
    \var{NodeTest}
  & \to & \var{KindTest}\,|\,\var{NameTest} & [78] \\
    \var{Literal}
  & \to & \var{NumericLiteral}\,|\,\var{StringLiteral} & [85] \\
    \var{VarName}
  & \to & \var{QName} & [88] \\
    \var{StringLiteral}
  & \to & \xquery{"$\cdots$"} & [144] \\
\end{array}
$
\caption{%
  Relevant \XQuery{} subset (source language).  Annotations
  in $[\cdot]$ refer to the grammar rules 
  in~\protect\cite[Appendix~A]{W3C:XQuery}.}
\label{grammar}
\end{figure}

Isolating the join graph implied by the input \XQuery{} expression lets
the relational database query optimizer face a problem known inside out
despite the source language not being \SQL{}: in essence, the join
graph isolation process emits a bundle of self-joins over the tabular
\XML{} document encoding connected by conjunctive equality and range
predicates. Most interestingly, we have found relational query
optimizers to be autonomously capable of translating these join graphs
into join trees that, effectively,
\begin{inparaenum}[(1)]
\item perform cost-based shuffling of the evaluation order of \XPath{} 
  location steps and predicates,
\item exploit \XPath{} axis reversal (\emph{e.g.}, trade
 \xquery{ancestor} for \xquery{descendant}), and
\item break up and stitch complex path expressions.
\end{inparaenum}
In recent years, all of these have been described as specific evaluation
and optimization techniques in the \XPath{} and \XQuery{} domain
\cite{HolisticTwigJoin,QueryOptimizationforXML,LookingForward}---here,
instead, they are the \emph{automatic result} of join tree planning
solely based on the availability of vanilla B-tree indexes and
associated statistics. The resulting plans fully exploit the relational
database kernel infrastructure, effectively turning the RDBMS into an
\XQuery{} processor that can perfectly cope with large \XML{} instances
(of size~100\,MB and beyond).

We plugged join graph isolation into
\Pathfinder{}\footnote{\url{http://www.pathfinder-xquery.org/}}---a
full-fledged compiler for the complete \XQuery{} language specification
targeting conventional relational database back-ends---and observed
significant query execution time improvements for popular \XQuery{}
benchmarks, \emph{e.g.}, XMark or the query section of TPoX
\cite{TPoX,XMark}.

\smallskip\noindent 
We start to explore this form of \XQuery{} join
graph isolation in Section~\ref{join-based-xquery-semantics} where we
review the compiler's algebraic target language, tabular \XML{} document
encodings, and join-based compilation rules for \XPath{} location steps,
nested \xquery{for} loops, and conditionals. The rewriting
procedure of Section~\ref{xquery-join-graph-isolation} then isolates the
join graphs buried in the initial compiled plans. Cast in terms of an
\SQL{} query, IBM\,DB2\,V9's relational query processor is able turn
these graphs into join trees which, effectively, implement a series of
otherwise \XQuery{}- and \XPath{}-specific optimizations. A further
quantitative experimental assessment demonstrates that DB2\,V9's
built-in \pureXML{} \XQuery{} processor currently faces a serious
challenger with its relational self if the latter is equipped with the
join graph-isolating compiler
(Section~\ref{db2-udb-xquery-wizard-builtin}).
Sections~\ref{more-related-work} and~\ref{work-in-flux} conclude this
paper with reviews of related efforts and work in flux.

\begin{table}
\centering\small
\caption{Table algebra dialect (compilation target language).}
\label{relational-operators}
\begin{tabular}{@{}ll@{}}
  \toprule
    \textbf{Operator}
  & \textbf{Semantics} 
  \\
    \cmidrule(r){1-1}\cmidrule(l){2-2}
    $\ser$
  & serialization point (plan root)
  \\
    $\pi_{\col{a}_1:\col{b}_1, \ndots, \col{a}_n:\col{b}_n}$
  & project onto columns $\col{b}_i$, rename $\col{b}_i$ into $\col{a}_i$
  \\
    $\select_p$
  & select rows that satisfy predicate $p$
  \\
    $\_ \Join_p \_$
  & join with predicate $p$
  \\
    $\_ \cross \_$
  & Cartesian product
  \\
    $\distinct$
  & eliminate duplicate rows
  \\
    $\attach_{\col{a}:c}$
  & attach column $\col{a}$ containing constant value $c$
  \\
    $\num_{\col{a}}$
  & attach arbitrary unique row id in column $\col{a}$
  \\
    $\rank_{\col{a}:\left<\col{b}_1,\ndots,\col{b}_n\right>}$
  & attach row rank in $\col{a}$ (in $\col{b}_1,\ndots,\col{b}_n$ order)
  \\
    $\doc$
  & \XML{} document encoding table
  \\
    $
       \begin{littbl}
         \begin{array}[t]{|c|c|}
           \colhd{\scriptstyle a} & \colhd{\scriptstyle b} \\
           \scriptstyle c_1 & \scriptstyle c_2 \\
           \hline
         \end{array} 
       \end{littbl}
    $
  & singleton literal table (with columns $\col{a}$, $\col{b}$)
  \\
\bottomrule
\end{tabular}
\end{table}

\section{Join-Based XQuery Semantics}
\label{join-based-xquery-semantics}

To prepare join graph isolation, the compiler translates the \XQuery{}
fragment of Fig.~\ref{grammar} into intermediate DAG-shaped plans over
the table algebra of Table~\ref{relational-operators}. This
particularly simple algebra dialect has been designed to match the
capabilities of \SQL{} query engines: operators consume tables (not
relations) and duplicate row elimination is explicit (in terms
of~$\distinct$). The row rank operator
$\rank_{\col{a}:\left<\col{b}_{1},\dots,\col{b}_{n}\right>}$ exactly
mimics \SQL:1999's
\sql{RANK()\,OVER\,(ORDER\,BY\,$\col{b}_1,\dots,\col{b}_n$)\,AS\,$\col{a}$}
and is primarily used to account for \XQuery{}'s pervasive sequence
order notion. The attach operator
$\attach_{\col{a}:c}(e)$ abbreviates $e \cross 
\begin{littbl}
  \begin{array}{|c|}
    \colhd{\scriptstyle a} \\
    \scriptstyle c \\
    \hline
  \end{array} 
\end{littbl}$\,,
where the right-hand side argument denotes a singleton literal table.
Operator $\ser$ marks the root of the plan DAG, delivering those rows 
that encode the resulting \XML{} node sequence. Below we will see how
the join operator $\Join$ assumes a central role in the translation of
\XPath{} location steps, \xquery{for} loops, and conditional
expressions.

\begin{figure}
  \centering\small
  \begin{minipage}[c]{0.37\linewidth}
  \begin{lstlisting}[basicstyle=\ttfamily,basewidth=0.475em]
<open_auction id="1">
 <initial>
  15
 </initial>
 <bidder>
  <time>18:43</time>
  <increase>
   4.20
  </increase>
 </bidder>
</open_auction>
  \end{lstlisting}
  \end{minipage}
  $
  \begin{littbl}
  \renewcommand{\arraystretch}{0.8}
    \begin{array}[c]{@{}|c|c|c|c|l|r|.|@{}}
      \colhd{pre} & \colhd{size} & \colhd{level} & \colhd{kind} &
      \colhd{name} & \colhd{value} & \colhd{data} 
      \\
      0  & 9 & 0 & \xquery{DOC}  & \xquery{au$\cdots$xml}  &              &
      \\
      1  & 8 & 1 & \xquery{ELEM} & \xquery{open\_$\cdots$} &              & 
      \\
      2  & 0 & 2 & \xquery{ATTR} & \xquery{id}            & \xquery{1}     & 1.0
      \\ 
      3  & 1 & 2 & \xquery{ELEM} & \xquery{initial}       & \xquery{15}    & 15.0
      \\
      4  & 0 & 3 & \xquery{TEXT} &                        & \xquery{15}    & 15.0
      \\
      5  & 4 & 2 & \xquery{ELEM} & \xquery{bidder}        &                &
      \\
      6  & 1 & 3 & \xquery{ELEM} & \xquery{time}          & \xquery{18:43} &
      \\
      7  & 0 & 4 & \xquery{TEXT} &                        & \xquery{18:43} &
      \\
      8  & 1 & 3 & \xquery{ELEM} & \xquery{incre$\cdots$}      & \xquery{4.20}  & 4.2
      \\
      9 &  0 & 4 & \xquery{TEXT} &                        & \xquery{4.20}  & 4.2
      \\
      \hline
    \end{array}
  \end{littbl}
  $
  \caption{Encoding of the infoset of \XML{} document \xquery{auction.xml}.
  Column  $\col{data}$ carries the nodes' typed decimal values.}
  \label{encoded-XML}
\end{figure}

\subsection{\XML{} Infoset Encoding}
An encoding of persistent \XML{} infosets is provided via the designated
table $\doc$. In principle, any schema-oblivious node-based encoding of
\XML{} nodes that admits the evaluation of \XPath{} node tests and axis
steps fits the bill (\emph{e.g.}, \emph{ORDPATH}~\cite{ORDPATH}). The
following uses one such row-based format in which, for each node $v$, key
column $\col{pre}$ holds $v$'s unique document order rank to form---together
with columns $\col{size}$ (number of nodes in subtree below $v$) and
$\col{level}$ (length of path from $v$ to its document root node)---an
encoding of the \XML{} tree structure (Fig.~\ref{encoded-XML} and
\cite{XPathAccel}). \XPath{} kind and name tests access columns $\col{kind}$
and $\col{name}$---multiple occurrences of value $\xquery{DOC}$ in column
$\col{kind}$ indicate that table $\doc$ hosts several trees, distinguishable
by their document URIs (in column $\col{name}$). For nodes with $\col{size}
\leqslant 1$, table $\doc$ supports value-based node access in terms of two
columns that carry the node's untyped string
value~\cite[\S\,3.5.2]{W3C:XQueryFS} and, if applicable, the result of a
cast to type \xquery{xs:decimal}\footnote{In the interest of space, we omit
a discussion of the numerous further \XML~Schema built-in data types.}
(columns $\col{value}$ and $\col{data}$, respectively). This tabular \XML{}
infoset representation may be efficiently populated (during a single parsing
pass over the \XML{} document text) and serialized again (via a table scan
in $\col{pre}$ order).

\begin{figure}[tht]
\centering\small
$
\begin{array}{@{}ll@{}}
  \toprule
  n & \var{kindt}(n) \\
  \cmidrule(r){1-1}\cmidrule(l){2-2}  
  \xquery{element($\_$,$\_$)}   & \col{kind} = \sql{ELEM}    \\
  \xquery{attribute($\_$,$\_$)} & \col{kind} = \sql{ATTR}    \\
  \xquery{text()}               & \col{kind} = \sql{TEXT}    \\[-1ex]
  \multicolumn{1}{c}{\vdots}    & \multicolumn{1}{c}{\vdots} \\[-0.5ex]
  \bottomrule
\end{array}
\quad\,\,
\begin{array}{@{}ll@{}}
  \toprule
  n & \var{namet}(n) \\
  \cmidrule(r){1-1}\cmidrule(l){2-2}  
  \xquery{element($t$,$\_$)}    & \col{name} = t    \\
  \xquery{attribute($t$,$\_$)}  & \col{name} = t    \\
  \xquery{text()}               & \var{true}        \\[-1ex]
  \multicolumn{1}{c}{\vdots}    & \multicolumn{1}{c}{\vdots} \\[-0.5ex]
  \bottomrule
\end{array}
$

\bigskip
$
\begin{array}{@{}ll@{}}
  \toprule
  \alpha & \var{axis}(\alpha) \\
  \cmidrule(r){1-1}\cmidrule(l){2-2}  
  \xquery{child}      & \col{pre}_{\circ} < \col{pre} \leqslant \col{pre}_{\circ} + \col{size}_{\circ}
    \wedge \col{level}_{\circ} + 1 = \col{level}    \\
  \xquery{descendant} & \col{pre}_{\circ} < \col{pre} \leqslant \col{pre}_{\circ} + \col{size}_{\circ} \\
  \xquery{ancestor}   & \col{pre} < \col{pre}_{\circ} \leqslant \col{pre} + \col{size} \\
  \xquery{following}  & \col{pre}_{\circ} + \col{size}_{\circ} < \col{pre} \\[-1ex]
  \multicolumn{1}{c}{\vdots} & \multicolumn{1}{c}{\vdots} \\[-0.5ex]
  \bottomrule
\end{array}
$
\caption{Predicates implementing the semantics of \XPath{} kind and
  name tests---expressed in sequence type syntax 
  \protect\cite[\S\,2.5.3]{W3C:XQuery}---and axes (excerpt).
  $\circ$ marks the properties of the context node(s).}
\label{kindtnamet}
\end{figure}

\subsection{\XPath{} Location Steps}
Further, this encoding has already been shown to admit
the efficient join-based evaluation of location steps
$\alpha\xquery{::}n$ along all 12~\XPath{} axes $\alpha$
\cite{XPathAccel}. While the structural node relationship expressed by
$\alpha$ maps into a conjunctive range join predicate
$\var{axis}(\alpha)$ over columns $\col{pre}$, $\col{size}$,
$\col{level}$, the step's kind and/or name test $n$ yields equality
predicates over $\col{kind}$ and $\col{name}$ (Fig.~\ref{kindtnamet}).
Consider the three-step
path $Q_{0}\!=\!\xquery{doc("auction.xml")/descendant::bidder/}$
\xquery{child::*/child::text()} over the document of 
Fig.~\ref{encoded-XML}.  To perform the final \xquery{child::text()}
step, which will have context elements
\xquery{time} and \xquery{increase}, the database system evaluates a
join between the document encoding and the step's context nodes
(the query yields the $\col{pre}$ ranks of the two resulting text nodes): 
$$
\mskip-2mu\pi_{%
  \begin{array}[c]{@{}l@{}}
      \scriptstyle
      \col{item}: \\[-1.2ex]
      \scriptstyle
      \col{pre}
   \end{array}}\!\biggl(\mskip-4mu\sigma_{%
  \begin{array}[c]{@{}l@{}}
      \scriptstyle
      \var{kindt}(\xquery{text()}) \mathbin{\wedge} \\[-1.2ex]
      \scriptstyle
      \var{namet}(\xquery{text()})
   \end{array}}\!(\doc) 
\mskip-6mu\underset{\var{axis}(\xquery{child})}{\Join}
\begin{littbl}
  \small
  \begin{array}[c]{|c|c|c|c|}
    \colhd{\smash{\col{pre}}_{\pmb{\circ}}} &
    \colhd{\col{size}_{\pmb{\circ}}} &
    \colhd{\col{level}_{\pmb{\circ}}} &
    \colhd{\!\cdots\!} \\
    6 &  1 &  3 & \\
    8 &  1 &  3 & \raisebox{.2ex}{$\smash{\vdots}$} \\
    \hline    
  \end{array}
\end{littbl}
\biggr) 
\mskip-3mu=\mskip-2mu
\begin{littbl}
  \small
  \begin{array}[c]{|c|}
    \colhd{\col{item}} \\
    7 \\
    9 \\
    \hline    
  \end{array}
\end{littbl}\,.
$$
With their ability to perform range scans, regular B-tree indexes,
built over table $\doc$, perfectly support this
style of location step evaluation 
\cite{XPathAccel}.  

\subsection{A Loop-Lifting \XQuery{} Compiler}
\label{loop-lifting-xquery-compiler}
                 
From \cite{XQueryOnSQLHosts} we adopt a view of the dynamic \XQuery{}
semantics, \emph{loop lifting}, that revolves around the \xquery{for} loop
as the core language construct. \emph{Any} subexpression $e$ is considered
to be iteratively evaluated inside its innermost enclosing \xquery{for}
loop.  For the \XQuery{} fragment of Fig.~\ref{grammar}, each iterated
evaluation of $e$ yields a (possibly empty) ordered sequence of nodes. To
reflect this, we compile $e$ into an algebraic plan that returns a ternary
table with schema $\col{iter}|\col{pos}|\col{item}$: a row $[i,p,v]$
indicates that, in iteration $i$, the evaluation of $e$ returned a sequence
containing a node with $\col{pre}$ rank $v$ at sequence position $p$.

The inference rules \ref{compile-Doc}, \ref{compile-Ddo},
\ref{compile-Step}, \ref{compile-If-Then}, \ref{compile-Comp},
\ref{compile-For}, and \ref{compile-Var} (taken
from~\cite{XQueryOnSQLHosts} and reproduced in
Appendix~\ref{inference-rules}) form a \emph{compositional} algebraic
compilation scheme for the \XQuery{} dialect of Fig.~\ref{grammar}.
The rule set expects to see the input query after \XQuery{}~Core
normalization: the enforcement of duplicate node removal and document
order after \XPath{} location steps (via the application of
\xquery{fs:distinct-doc-order($\cdot$)}, abbreviated to
\xquery{fs:ddo($\cdot$)} in the following) and the computation of
effective Boolean values in conditionals (via
\xquery{fn:boolean($\cdot$)}) is explicit \cite[\S\,4.2.1 and
\S\,3.4.3]{W3C:XQueryFS}.
                 
\subsection{The Compositionality Threat}
\label{impact-of-compositionality}

To obtain an impression of typical plan features, we compile
\begin{equation*}
\begin{BVerbatim}[baseline=c,commandchars=\\\{\}]
       doc("auction.xml")
/descendant::open\_auction[bidder]\enskip\text.
\end{BVerbatim}
\tag{$Q_1$}
\label{Q1}
\end{equation*}
After \XQuery{}~Core normalization, this query reads
\begin{equation*}
\begin{BVerbatim}[baseline=t,commandchars=\\\{\}]
for \vardollar{}x in fs:ddo(doc("auction.xml") 
                 /descendant::open_auction)
return if (fn:boolean(fs:ddo(\vardollar{}x/child::bidder)))
       then \vardollar{}x else ()\enskip\text.
\end{BVerbatim}
\end{equation*}
Fig.~\ref{plan-Q1} shows the initial plan for~\ref{Q1}. Since the
inference rules of Fig.~\ref{rules} implement a \emph{fully
compositional} compilation scheme, we can readily identify how the
subexpressions of~\ref{Q1} contribute to the overall plan (to this end,
observe the gray plan sections all of which yield tables with columns
$\col{iter}|\col{pos}|\col{item}$). \XQuery{} is a functional
expression-oriented language in which subexpressions are stacked upon
each other to form complex queries. The tall plan profile with its
stacked sections---reaching from a single instance of table $\doc$
(serving all node references) to the serialization
point~$\ser$\,---directly reflects this orthogonal nesting of
expressions.

\smallskip\noindent
Note, though, how this artifact of both, compositional language and
compilation scheme, leads to plans whose shapes differ considerably from
the ideal \emph{join graph $+$ plan tail} we have identified earlier.
Instead, join operators occur in sections distributed all
over the plan. A similar distribution can be observed for the
blocking operators $\distinct$ and $\rank$ (duplicate elimination and
row ranking). This is quite unlike the algebraic plans produced by \SQL{}
\SELECT-\FROM-\WHERE{} block compilation.

The omnipresence of blocking operators obstructs join operator movement
and planning and leads industrial-strength optimizers, \emph{e.g.},
IBM\,DB2\,UDB\,V9, to execute the plan in stages that read and then
again materialize temporary tables. In the following we will therefore
follow a different route and instead \emph{reshape} the plan into a join
graph that becomes subject to efficient one-shot execution by the \SQL{}
database back-end. (Section~\ref{db2-udb-xquery-wizard-builtin} will
show that join graph isolation for~\ref{Q1} improves the
evaluation time by a factor of~5.)

\section{XQuery Join Graph Isolation}
\label{xquery-join-graph-isolation}

In a nutshell, join graph isolation pursues a strategy that moves the
blocking operators ($\rank$ and $\distinct$) into plan tail positions
and, at the same time, pushes join operators down into the plan. This
rewriting process will isolate a plan section, the \emph{join graph},
that is populated with references to the infoset encoding table $\doc$,
joins, and further pipelineable operators, like projection, selection, and
column attachment ($\pi$, $\sigma$, $\attach$).

The ultimate goal is to form a new DAG that may readily be translated
into a \emph{single} \SELECT-\DISTINCT-\FROM-\WHERE-\ORDERBY{} block in which
\begin{compactenum}[(1)]
\item the \FROM{} clause lists the required $\doc$ instances,
\item the \WHERE{} clause specifies a conjunctive self-join predicate
  over $\doc$, reflecting the semantics of \XPath{} location steps
  and predicates, and 
\item the \SELECT-\DISTINCT{} and \ORDERBY{} clauses represent the plan tail. 
\end{compactenum}  

\subsection{Plan Property Inference}
We account for the unusual tall shape and substantial size (of the order of
100~operators and beyond for typical bench\-mark-type queries) of the
initial plan DAGs by a \emph{peep-hole}-style rewriting process.  For all
operators~$\varoast$, a property inference collects relevant information
about the plan vicinity of~$\varoast$.  The applicability of a rewriting
step may then decided by inspection of the properties of a single operator
(and its closer neighborhood) at a time.
Tables~\ref{tab:icols-property}--\ref{tab:distinct-property} define these
properties and their inference in an operator-by-operator fashion.  We rely
on auxiliary function $\var{cols}(\cdot)$ that can determine the columns
used in a predicate (\emph{e.g.},
$\var{cols}\mskip-4mu\left(\col{pre}_{\circ} + \col{size}_{\circ} < \col{pre}\right) = 
\{\col{pre}_{\circ},\col{size}_{\circ},\col{pre}\}$) as well as the
columns in the output table of a given plan fragment (\emph{e.g.},
$\var{cols}\mskip-4mu\left(\attach_{\col{iter}:1}\mskip-4mu\left(%
  \begin{littbl}
   \begin{array}{|c|}
     \colhd{\scriptstyle pos} \\
     \scriptstyle 1 \\
     \hline
   \end{array} 
  \end{littbl}
\right)\mskip-4mu\right) = \{\col{iter},\col{pos}\}$).
Furthermore we use $\Rrightarrow$
to denote the reachability relation of this DAG (\emph{e.g.}, $\varoast
\Rrightarrow \ser$ for any operator $\varoast$ in the plan).

\begin{figure}
  \footnotesize
  \vfill
  \hspace*{-5mm}\rlap{%
\begin{tikzpicture}[scale=0.5,>=to,join=miter,y=0.28mm,x=0.45mm]
  \tikzstyle{blob}=[rounded corners=20pt,draw=black!40,fill=black!7]                                                                                              \draw [blob] (200,40) -- (370,-10) -- (360,120) -- (280,215) --                              (210,180) -- cycle;                                                \node (doc-auction-xml) at (125,145) {                                            $\begin{array}{@{}r@{}}                                                           \xquery{doc(\textquotedblright{}auction.xml\textquotedblright{})} \\            \eqref{compile-Doc}                                                             \end{array}$};                                                                \draw [densely dotted] (doc-auction-xml.east) -- +(30,-20);                                                                                                     \draw [blob] (190,490) -- (140,440) -- (110,200) -- (340,215) --                             (310,320) -- cycle;                                                \node (descendant-open-auction) at (55,470) {                                     $\begin{array}{@{}r@{}}                                                           \xquery{descendant::open\_auction} \\                                           \eqref{compile-Step}                                                            \end{array}$};                                                                \draw [densely dotted] (descendant-open-auction.south east) -- +(10,-20);                                                                                       \draw [blob] (130,1065) -- (70,1000) -- (-40,780) -- (180,790) --                            (170,1000) -- cycle;                                               \node (child-bidder) at (15,1010) {                                               $\begin{array}{@{}r@{}}                                                           \xquery{child::bidder} \\                                                       \eqref{compile-Step}                                                            \end{array}$};                                                                \draw [densely dotted] (child-bidder) -- +(50,-50);                                                                                                             \draw [blob] (210,1380) -- (130,1300) -- (90,1200) -- (170,1210) --                          (210,1270) -- cycle;                                               \node (if-then) at (55,1360) {                                                    $\begin{array}{@{}r@{}}                                                           \xquery{if ($\cdots$) then \$x else ()} \\                                      \eqref{compile-If-Then}                                                         \end{array}$};                                                                \draw [densely dotted] (if-then.south east) -- +(10,-20);                                                                                                       \draw [blob] (160,630) -- (220,640) -- (280,700) -- (335,1300) --                            (300,1400) -- (280,1505) -- (200,1490) --                                       (140,1500) {[rounded corners=5pt] -- (207,1370) --                              (220,1280)} -- (240,1000) --                                                    (180,790) -- (140,720) -- cycle;                                   \node (for-return) at (55,1480) {                                                 $\begin{array}{@{}r@{}}                                                           \xquery{for \$x in $\cdots$ return $\cdots$} \\                                 \eqref{compile-For}                                                             \end{array}$};                                                                \draw [densely dotted] (for-return.east) -- +(30,-20);                                                                                                          \draw [blob][rotate=0] (190,555) ellipse (40 and 70);                           \node (fs-ddo1) at (270,600) {                                                    $\begin{array}{@{}l@{}}                                                           \xquery{fs:ddo($\cdot$)} \\                                                     \eqref{compile-Ddo}                                                             \end{array}$};                                                                \draw [densely dotted] (fs-ddo1) -- +(-50,-50);                                                                                                                 \draw [blob][rotate=0] (130,1125) ellipse (40 and 70);                          \node (fs-ddo2) at (50,1170) {                                                    $\begin{array}{@{}r@{}}                                                           \xquery{fs:ddo($\cdot$)} \\                                                     \eqref{compile-Ddo}                                                             \end{array}$};                                                                \draw [densely dotted] (fs-ddo2) -- +(50,-50);                                %
\node (ser1_) at (224,1527) [draw,draw=none] {$\relax$};
  \node (for1_1) at (224,1479) [draw,draw=none] {$\pi_{\col{iter}:\col{outer},\col{pos}:\col{pos_1},\col{item}}$};
  \node (doc1_) at (145,16) [draw,draw=none] {$\doc$};
  \node (loop1_) at (347,16) [draw,draw=none] {$\begin{littbl}                               \begin{array}{|c|}                         \colhd{\scriptstyle iter}                  \\                                       {\scriptstyle 1}                           \\                                       \hline                                     \end{array}                              \end{littbl}$};
  \node (for1_2) at (224,1428) [draw,draw=none] {$\rank_{\col{pos1}:\langle\col{sort},\col{pos}\rangle}$};
  \node (for1_3) at (224,1377) [draw,draw=none] {$\underset{\col{iter}=\col{inner}}{\Join}$};
  \node (for1_4) at (189,748) [draw,draw=none] {$\attach_{\col{pos}:1}$};
  \node (for1_5) at (189,698) [draw,draw=none] {$\pi_{\col{iter}:\col{inner},\col{item}}$};
  \node (for1_6) at (274,1273) [draw,draw=none] {$\pi_{\renewcommand{\arraystretch}{0.1}                          \begin{array}{@{}l@{}}                                       \scriptstyle                                               \col{outer}:\col{iter},\col{inner},\col{sort}:\col{pos}                      \end{array}}$};
  \node (for1_7) at (189,648) [draw,draw=none] {$\num_{\col{inner}}$};
  \node (ifthen1_1) at (189,1325) [draw,draw=none] {$\underset{\col{iter}_1=\col{iter}}{\Join}$};
  \node (ddo1_1) at (189,598) [draw,draw=none] {$\rank_{\col{pos}:\langle\col{item}\rangle}$};
  \node (ddo2o1_1) at (131,1171) [draw,draw=none] {$\rank_{\col{pos}:\langle\col{item}\rangle}$};
  \node (ddo2o1_2) at (129,1123) [draw,draw=none] {$\distinct$};
  \node (ddo2o1_3) at (126,1075) [draw,draw=none] {$\pi_{\col{iter},\col{item}}$};
  \node (child1_1) at (125,1024) [draw,draw=none] {$\rank_{\col{pos}:\langle\col{item}\rangle}$};
  \node (child1_2) at (118,973) [draw,draw=none] {$\pi_{\col{iter},\col{item}:\col{pre}}$};
  \node (child1_3) at (110,918) [draw,draw=none] {$\underset{                                                           \renewcommand{\arraystretch}{0.1}                                                   \begin{array}{@{}c@{}}                                                                \scriptstyle                                                                        \col{pre}_{\circ} < \col{pre} \leqslant \col{pre}_{\circ} + \col{size}_{\circ} \wedge {}     \\                                                                                \scriptstyle                                                                        \col{level}_{\circ} + 1 = \col{level}                                             \end{array}}{\Join}$};
  \node (child1_4) at (33,802) [draw,draw=none] {$\select_{\renewcommand{\arraystretch}{0.1}                                  \begin{array}{@{}l@{}}                   													  \scriptstyle                                                       (\col{kind} = \kind{ELEM}) \wedge {}                               \\                                   													  \scriptstyle                                                       (\col{name} = \xquery{'bidder'})                                 \end{array}}$};
  \node (child1_5) at (110,860) [draw,draw=none] {$\pi_{\renewcommand{\arraystretch}{0.1}                                  \begin{array}{@{}l@{}}                                               \scriptstyle                                                       \col{iter},\col{pre}_{\circ}:\col{pre},                            \\                                                               \scriptstyle                                                       \col{size}_{\circ}:\col{size},\col{level}_{\circ}:\col{level}                       \end{array}}$};
  \node (child1_6) at (110,802) [draw,draw=none] {$\underset{\col{pre}=\col{item}}{\Join}$};
  \node (ifthen1_2) at (142,1273) [draw,draw=none] {$\distinct$};
  \node (ifthen1_3) at (132,1222) [draw,draw=none] {$\pi_{\col{iter}_1:\col{iter}}$};
  \node (ddo1_2) at (189,550) [draw,draw=none] {$\distinct$};
  \node (ddo1_3) at (189,502) [draw,draw=none] {$\pi_{\col{iter},\col{item}}$};
  \node (descendant1_1) at (189,451) [draw,draw=none] {$\rank_{\col{pos}:\langle\col{item}\rangle}$};
  \node (descendant1_2) at (189,400) [draw,draw=none] {$\pi_{\col{iter},\col{item}:\col{pre}}$};
  \node (descendant1_3) at (189,348) [draw,draw=none] {$\underset{\col{pre}_{\circ} < \col{pre} \leqslant \col{pre}_{\circ} + \col{size}_{\circ}}{\Join}$};
  \node (descendant1_4) at (179,235) [draw,draw=none] {$\select_{\renewcommand{\arraystretch}{0.1}                                              \begin{array}{@{}l@{}}                          																 \scriptstyle                                                                   (\col{kind} = \kind{ELEM}) \wedge {}                                           \\                                          																 \scriptstyle                                  																 (\col{name} = \xquery{'open\_auction'})                                      \end{array}}$};
  \node (descendant1_5) at (264,293) [draw,draw=none] {$\pi_{\renewcommand{\arraystretch}{0.1}                                       \begin{array}{@{}l@{}}                       														 \scriptstyle                                                            \col{iter},\col{pre}_{\circ}:\col{pre},                                 \\                                       														 \scriptstyle                                                            \col{size}_{\circ}:\col{size},\col{level}_{\circ}:\col{level}                            \end{array}}$};
  \node (descendant1_6) at (267,235) [draw,draw=none] {$\underset{\col{pre}=\col{item}}{\Join}$};
  \node (doc1_1) at (268,180) [draw,draw=none] {$\pi_{\col{iter},\col{pos},\col{item}:\col{pre}}$};
  \node (doc1_2) at (269,132) [draw,draw=none] {$\cross$};
  \node (doc1_3) at (269,80) [draw,draw=none] {$\select_{\renewcommand{\arraystretch}{0.1}                                 \begin{array}{@{}l@{}}                                            _{(\col{kind} = \kind{DOC}) \wedge {}}                            \\                                                              _{(\col{name} = \xquery{'auction.xml'})}                         \end{array}}$};
  \node (doc1_4) at (347,80) [draw,draw=none] {$\attach_{\col{pos}:1}$};
  \draw [<-,o-] (ser1_) ..controls (224,1504) and (224,1493)  .. (for1_1);
  \draw [<-] (for1_1) ..controls (224,1453) and (224,1442)  .. (for1_2);
  \draw [<-] (for1_2) ..controls (224,1402) and (224,1391)  .. (for1_3);
  \draw [<-] (for1_3) ..controls (207,1352) and (199,1341)  .. (ifthen1_1);
  \draw [<-] (for1_3) ..controls (243,1338) and (262,1296)  .. (for1_6);
  \draw [<-] (for1_6) ..controls (269,1235) and (265,1200)  .. (265,1171) .. controls (265,1171) and (265,1171)  .. (265,748) .. controls (265,704) and (218,667)  .. (for1_7);
  \draw [<-] (for1_4) ..controls (189,723) and (189,712)  .. (for1_5);
  \draw [<-] (for1_5) ..controls (189,671) and (189,660)  .. (for1_7);
  \draw [<-] (for1_7) ..controls (189,623) and (189,612)  .. (ddo1_1);
  \draw [<-] (ddo2o1_1) ..controls (130,1144) and (129,1133)  .. (ddo2o1_2);
  \draw [<-] (ddo2o1_2) ..controls (127,1100) and (127,1089)  .. (ddo2o1_3);
  \draw [<-] (ddo2o1_3) ..controls (126,1049) and (126,1038)  .. (child1_1);
  \draw [<-] (child1_1) ..controls (121,998) and (120,987)  .. (child1_2);
  \draw [<-] (child1_2) ..controls (114,947) and (113,937)  .. (child1_3);
  \draw [<-] (child1_3.west) ..controls (24,905) and (28,839)  .. (child1_4);
  \draw [<-] (child1_3) ..controls (110,887) and (110,877)  .. (child1_5);
  \draw [<-] (child1_4) ..controls (36,760) and (37,727)  .. (37,698) .. controls (37,698) and (37,698)  .. (37,132) .. controls (37,68) and (124,25)  .. (doc1_);
  \draw [<-] (child1_5) ..controls (110,829) and (110,817)  .. (child1_6);
  \draw [<-] (child1_6) ..controls (104,763) and (99,728)  .. (99,698) .. controls (99,698) and (99,698)  .. (99,132) .. controls (99,84) and (132,34)  .. (doc1_);
  \draw [<-] (child1_6) ..controls (146,778) and (168,762)  .. (for1_4);
  \draw [<-] (ifthen1_1) ..controls (165,1298) and (151,1283)  .. (ifthen1_2);
  \draw [<-] (ifthen1_2) ..controls (137,1249) and (134,1237)  .. (ifthen1_3);
  \draw [<-] (ifthen1_1) ..controls (189,1286) and (189,1251)  .. (189,1222) .. controls (189,1222) and (189,1222)  .. (189,860) .. controls (189,819) and (189,771)  .. (for1_4);
  \draw [<-] (ifthen1_3) ..controls (132,1196) and (132,1185)  .. (ddo2o1_1);
  \draw [<-] (ddo1_1) ..controls (189,571) and (189,560)  .. (ddo1_2);
  \draw [<-] (ddo1_2) ..controls (189,527) and (189,516)  .. (ddo1_3);
  \draw [<-] (ddo1_3) ..controls (189,476) and (189,465)  .. (descendant1_1);
  \draw [<-] (descendant1_1) ..controls (189,425) and (189,414)  .. (descendant1_2);
  \draw [<-] (descendant1_2) ..controls (189,374) and (189,363)  .. (descendant1_3);
  \draw [<-] (descendant1_3) ..controls (185,306) and (182,265)  .. (descendant1_4);
  \draw [<-] (descendant1_3) ..controls (221,324) and (239,312)  .. (descendant1_5);
  \draw [<-] (descendant1_4) ..controls (169,165) and (150,43)  .. (doc1_);
  \draw [<-] (descendant1_5) ..controls (266,262) and (266,250)  .. (descendant1_6);
  \draw [<-] (descendant1_6) ..controls (240,212) and (229,200)  .. (220,187) .. controls (188,139) and (194,120)  .. (171,69) .. controls (162,50) and (151,28)  .. (doc1_);
  \draw [<-] (descendant1_6) ..controls (267,207) and (267,195)  .. (doc1_1);
  \draw [<-] (doc1_1) ..controls (268,153) and (269,142)  .. (doc1_2);
  \draw [<-] (doc1_2) ..controls (269,109) and (269,98)  .. (doc1_3);
  \draw [<-] (doc1_2) ..controls (296,114) and (324,96)  .. (doc1_4);
  \draw [<-] (doc1_3) ..controls (208,48) and (161,24)  .. (doc1_);
  \draw [<-] (doc1_4) ..controls (347,54) and (347,41)  .. (loop1_);
\end{tikzpicture}
}
  \vfill
  \caption{%
    Initial stacked plan for~\ref{Q1}. The gray plan sections indicate
    input \XQuery{} subexpressions and applied compilation rules.}
  \label{plan-Q1}  
\end{figure}

\begin{table*}
\centering\footnotesize
\begin{minipage}[t]{0.55\linewidth}
\centering  
\caption{Top-down inference of property~$\icols$ for the input(s)
  of operator~$\varoast$.}
\label{tab:icols-property}
\vspace{-2ex}
\begin{tabular}[t]{@{}ll@{}}
  \toprule
    \textbf{Operator} $\varoast$
  & \textbf{Inferred property \icols{} of input(s) of} $\varoast$
  \\
    \cmidrule(r){1-1}\cmidrule(l){2-2}
    $\ser(e)$
  & $e.\icols \is 
       \{\col{pos},\col{item}\}$ 
  \\
    $\pi_{\col{a}_1:\col{b}_1, \ndots, \col{a}_n:\col{b}_n}\mskip-3mu(e)$
  & $e.\icols \is e.\icols \cup \{ \col{b}_i \,|\, \col{a}_i \in (\icols \cap \{ \col{a}_1,\ndots,\col{a}_n\})\}$
  \\
    $\select_p(e)$
  & $e.\icols \is e.\icols \cup \icols \cup \var{cols}(p)$
  \\
    $e_1 \Join_p e_2$
  & $e_{1,2}.\icols \is e_{1,2}.\icols \cup ((\icols \cup \var{cols}(p)) \cap \var{cols}(e_{1,2}))$
  \\
    $e_1 \cross e_2$
  & $e_{1,2}.\icols \is e_{1,2}.\icols \cup (\icols \cap \var{cols}(e_{1,2}))$
  \\
    $\distinct(e)$
  & $e.\icols \is e.\icols \cup \icols$
  \\
    $\attach_{\col{a}:c}(e)$
  & $e.\icols \is e.\icols \cup (\icols \setminus \{ \col{a} \})$
  \\
    $\num_{\col{a}}(e)$
  & $e.\icols \is e.\icols \cup (\icols \setminus \{ \col{a} \})$
  \\
    $\rank_{\col{a}:\left<\col{b}_1,\ndots,\col{b}_n\right>}(e)$
  & $e.\icols \is e.\icols \cup (\icols \setminus \{ \col{a} \}) \cup
    \{ \col{b}_1, \ndots, \col{b}_n \}$
  \\
    $\doc$
  & $-$
  \\
    $
       \begin{littbl}
         \begin{array}[t]{|c|c|}
           \colhd{\scriptstyle a} & \colhd{\scriptstyle b} \\
           \scriptstyle c_1 & \scriptstyle c_2 \\
           \hline
         \end{array} 
       \end{littbl}
    $
  & $-$
  \\
\bottomrule
\end{tabular}
\end{minipage}
\hfill
\begin{minipage}[t]{0.44\linewidth}
\centering
\caption{Bottom-up inference of property~\const{} for 
  operator~$\varoast$.}
\label{tab:const-property}
\vspace{-2ex}
\begin{tabular}[t]{@{}ll@{}}
  \toprule
    \textbf{Operator $\varoast$}
  & \textbf{Inferred property \const{} of $\varoast$}
  \\
    \cmidrule(r){1-1}\cmidrule(l){2-2}
    $\ser(e)$
  & $\const \is e.\const$
  \\
    $\pi_{\col{a}_1:\col{b}_1, \ndots, \col{a}_n:\col{b}_n}\mskip-3mu(e)$
  & $\const \is \{ \col{a}_i = c \,|\, (\col{b}_i = c) \in e.\const \}$
  \\
    $\select_p(e)$
  & $\const \is e.\const$
  \\
    $e_1 \Join_p e_2$
  & $\const \is e_1.\const \cup e_2.\const$
  \\
    $e_1 \cross e_2$
  & $\const \is e_1.\const \cup e_2.\const$
  \\
    $\distinct(e)$
  & $\const \is e.\const$
  \\
    $\attach_{\col{a}:c}(e)$
  & $\const \is e.\const \union \{ \col{a} = c \}$
  \\
    $\num_{\col{a}}(e)$
  & $\const \is e.\const$
  \\
    $\rank_{\col{a}:\left<\col{b}_1,\ndots,\col{b}_n\right>}(e)$
  & $\const \is e.\const$
  \\
    $\doc$
  & $\const \is \varnothing$
  \\
    $
     \begin{littbl}
       \begin{array}[t]{|c|c|}
         \colhd{\scriptstyle a} & \colhd{\scriptstyle b} \\
         \scriptstyle c_1 & \scriptstyle c_2 \\
         \hline
       \end{array} 
     \end{littbl}
    $
  & $\const \is \{ \col{a} = c_1, \col{b} = c_2\}$
  \\
\bottomrule
\end{tabular}
\end{minipage}
\end{table*}

\begin{table*}
\centering\footnotesize
\begin{minipage}[t]{0.64\linewidth}
\centering
\caption{Bottom-up inference of property~\key{} for operator~$\varoast$.}
\label{tab:key-property}
\vspace{-2ex}
\begin{tabular}[t]{@{}ll@{}}
  \toprule
    \textbf{Operator $\varoast$}
  & \textbf{Inferred property \key{} of $\varoast$}
          \\
    \cmidrule(r){1-1}\cmidrule(l){2-2}
    $\ser(e)$
  & $\key \is e.\key$
  \\
    $\pi_{\col{a}_1{:}\col{b}_1, \ndots, \col{a}_n{:}\col{b}_n}\mskip-3mu(e)$
  & $\key \is
    \{ \{ \col{a}_i \,|\, \col{b}_i \in k \} \,|\, k \in e.\key, k \subseteq \{ \col{b}_1,\ndots,\col{b}_n \} \}$
  \\
    $\select_p(e)$
  & $\key \is e.\key$
  \\
    $e_1 \Join_{\col{a} = \col{b}} e_2$
  & $\key
    \begin{array}[t]{@{\mskip5mu}c@{\mskip5mu}l@{}}
      \mathbin{\is} & \{ k_1 \,|\,
      \{\col{b}\} \in e_2.\key, k_1 \in e_1.\key \} \cup
      \{ k_2 \,|\, 
      \{\col{a}\} \in e_1.\key, k_2 \in e_2.\key \} 
      \\
      \mathbin{\cup} &
      \{ (k_1 \setminus \{\col{a}\}) \cup k_2 \,|\,
      \{\col{b}\} \in e_2.\key, k_1 \in e_1.\key, k_2 \in e_2.\key \} 
      \\
      \mathbin{\cup} &
      \{ k_1 \cup (k_2 \setminus \{\col{b}\})\,|\,
      \{\col{a}\} \in e_1.\key, k_1 \in e_1.\key, k_2 \in e_2.\key \} 
    \end{array}
    $  
  \\
    $e_1 \Join_p e_2$
  & $\key \is \{ k_1 \cup k_2 \,|\, k_1 \in e_1.\key, k_2 \in e_2.\key \}$
  \\
    $e_1 \cross e_2$
  & $\key \is \{ k_1 \cup k_2 \,|\, k_1 \in e_1.\key, k_2 \in e_2.\key \}$
  \\
    $\distinct(e)$
  & $\key \is e.\key \cup \{ \var{cols}(e)\}$
  \\
    $\attach_{\col{a}:c}(e)$
  & $\key \is e.\key$
  \\
    $\num_{\col{a}}(e)$
  & $\key \is e.\key \cup \{ \{\col{a}\}\}$
  \\
    $\rank_{\col{a}:\left<\col{b}_1,\ndots,\col{b}_n\right>}(e)$
  & $\key \is e.\key \cup 
    \{ \{ \col{a} \} \cup (k \setminus \{ \col{b}_1,\ndots,\col{b}_n \}) \,|\, k \in e.\key, 
    k \cap \{ \col{b}_1,\ndots,\col{b}_n\} \neq \varnothing \}$
  \\
    $\doc$
  & $\key \is \{ \{\col{pre}\}\}$
  \\
    $
       \begin{littbl}
         \begin{array}[t]{|c|c|}
           \colhd{\scriptstyle a} & \colhd{\scriptstyle b} \\
           \scriptstyle c_1 & \scriptstyle c_2 \\
           \hline
         \end{array} 
       \end{littbl}
    $
  & $\key \is \{ \{\col{a}\}, \{\col{b}\}, \{\col{a},\col{b}\} \}$
  \\
\bottomrule
\end{tabular}
\end{minipage}
\hfill
\begin{minipage}[t]{0.32\linewidth}
\centering
\caption{Top-down inference of \mbox{Boolean property} \set{} for the input(s)
  of operator~$\varoast$.}%
\label{tab:distinct-property}%
\vspace{-1ex}
\begin{tabular}[t]{@{}ll@{}}
  \\[-1.2ex]
  \toprule
    \textbf{Operator $\varoast$}
  & \textbf{Inferred property \set{}} 
  \\ 
  & \textbf{of the input(s) of $\varoast$}
  \\
    \cmidrule(r){1-1}\cmidrule(l){2-2}
    $\ser(e)$
  & $e.\set \is \var{false}$
  \\
    $\pi_{\col{a}_1:\col{b}_1,\ndots,\col{a}_n:\col{b}_n}\mskip-3mu(e)$
  & $e.\set \is e.\set \wedge \set$
  \\
    $\select_p(e)$
  & $e.\set \is e.\set \wedge \set$
  \\
    $e_1 \Join_p e_2$ 
  & $e_{1,2}.\set \is e_{1,2}.\set \wedge \set$
  \\
    $e_1 \cross e_2$
  & $e_{1,2}.\set \is e_{1,2}.\set \wedge \set$
  \\
    $\distinct(e)$
  & $e.\set \is e.\set \wedge \var{true}$
  \\
    $\attach_{\col{a}:c}(e)$
  & $e.\set \is e.\set \wedge \set$
  \\
    $\num_{\col{a}}(e)$
  & $e.\set \is e.\set \wedge \set$
  \\
    $\rank_{\col{a}:\left<\col{b}_1,\ndots,\col{b}_n\right>}(e)$
  & $e.\set \is e.\set \wedge \set$
  \\
    $\doc$
  & $-$
  \\
    $
     \begin{littbl}
       \begin{array}[t]{|c|c|}
         \colhd{\scriptstyle a} & \colhd{\scriptstyle b} \\
         \scriptstyle c_1 & \scriptstyle c_2 \\
         \hline
       \end{array} 
     \end{littbl}
    $
  & $-$
  \\
\bottomrule
\end{tabular}
\end{minipage}
\end{table*}

\smallskip\noindent
\textbf{icols} This property records the set of input columns
strictly required to evaluate $\varoast$ and its upstream plan. At the
plan root $\ser$, the property is seeded with the set $\{\col{pos},
\col{item}\}$, the two columns required to represent and serialize the
resulting \XML{} node sequence. The $\icols$ column set is inferred
top-down and accumulated whenever the DAG-walking inference enters a
node more than once.

\noindent
\textbf{const} A set with elements of the form $\col{a} = c$,
indicating that \emph{all} rows in the table output by $\varoast$ hold
value $c$ in column $\col{a}$. Seeded at the plan leaves (instances of
$\doc$ or literal tables) and inferred bottom-up.

\noindent
\textbf{key} The set of candidate keys generated by $\varoast$ (bottom-up).

\noindent
\textbf{set} 
Boolean property $\set$ communicates whether the output rows of $\varoast$
will undergo duplicate elimination in the upstream plan.  Inferred top-down
($\set$ is initialized to $\var{true}$ for all operators but $\ser$).   

\subsection{Isolating Plan Tail and Join Graph}
The isolation process is defined by three subgoals 
\circled{\scriptsize\raisebox{1pt}{$\rank$}},
\circled{\scriptsize$\distinct$}, and \circled{\scriptsize\raisebox{0.75pt}{$\Join$}}
(described below), attained through a sequence of goal-directed
applications of the rewrite rules in Fig.~\ref{rewrite-rules}. Note
how the rules' premises inspect the inferred plan properties, as
described above. In addition to these three main goals, ``house
cleaning'' is performed by rules defined to simplify or remove operators
(Rules~(\ref{rewrite-1}--\ref{rewrite-5}), \eqref{rewrite-7}, and
\eqref{rewrite-13}). Otherwise, the subgoals either strictly move
$\rank$ and $\distinct$ towards the plan tail or push $\Join$ down into
the plan.  Progress and termination thus is guaranteed.
Goal~\circled{\scriptsize\raisebox{1pt}{$\rank$}} is pursued first.

\smallskip\noindent
\circled{\scriptsize\raisebox{1pt}{$\rank$}}~\textbf{%
Establish a single $\rank$ operator in the plan tail.}                      
A \emph{single} instance of the row ranking operator $\rank$ suffices to
correctly implement the sequence and document order requirements of
the overall plan.  To this end, Rule~\eqref{rewrite-12} trades a $\rank$
for a projection if $\rank$ ranks over a single column.  For the compilation
Rules~\ref{compile-Ddo} and~\ref{compile-Step}, which introduce 
row rankings of this form ($\rank_{\col{pos}:\left<\col{item}\right>}$),
this effectively means that document order determines sequence 
order---which is indeed the case for the result of \XPath{} location steps and
\xquery{fs:ddo($\cdot$)}.  All other instances of $\rank$ 
($\rank_{\col{pos}_{1}:\left<\col{sort},\col{pos}\right>}$, introduced
by \ref{compile-For}) are moved towards
the plan tail via Rules~\eqref{rewrite-14}--\eqref{rewrite-17}.  The
premises of Rules~\eqref{rewrite-14} and~\eqref{rewrite-15} are no obstacle
here: for the \XQuery{} fragment of Fig.~\ref{grammar}, the compiler
does not emit predicates over sequence positions (column~$\col{pos}$).  
Once arrived in the plan
tail, Rule~\eqref{rewrite-17} splices the ranking criteria of adjacent
$\rank$ operators.  Rule~\eqref{rewrite-2} finally removes all but the 
topmost instance of $\rank$.

\smallskip\noindent
\circled{\scriptsize$\distinct$}+\circled{\scriptsize\raisebox{0.75pt}{$\Join$}} 
\textbf{Establish a single $\distinct$ operator in the plan tail +
join push-down and removal.} Duplicate
elimination relocation and join push-down and removal are intertwined.
Fig.~\ref{join-pushdown} illustrates the stages of this process 
(the~\tikz[baseline=-2pt]    
   \tikzstyle{plan}=[circle,inner sep=0.5pt,
                     draw=black!40,fill=black!7]
   \node[plan] (P) {\phantom{3}};
represent plan sections much like in Fig.~\ref{plan-Q1}). These subgoals
target and ultimately delete the equi-joins introduced by the compilation 
Rules~\ref{compile-If-Then} and~\ref{compile-For} (the latter is in 
focus here). 

\begin{figure*}
\centering\small
\mprset{flushleft}
$$
\begin{array}{@{}c@{}}
\inferrule
  {\col{a} \notin \icols
  }{
   \num_{\col{a}}(q) \to q
  }(\lbl{rewrite}{1}{rewrite-1})
\qquad
\inferrule
  {\col{a} \notin \icols
  }{
   \rank_{\col{a}:\left<\col{b}_1,\dots,\col{b}_n\right>}(q) \to q
  }(\lbl{rewrite}{2}{rewrite-2})
\qquad
\inferrule
  {\col{a} \notin \icols
  }{
   \attach_{\col{a}:c}(q) \to q
  }(\lbl{rewrite}{3}{rewrite-3})
\qquad
\inferrule
  {\{\col{a}_1,\dots,\col{a}_n\} \cap \icols \neq \varnothing
  }{
   \pi_{\col{a}_1,\dots,\col{a}_n}(q) \to 
   \pi_{\{\col{a}_1,\dots,\col{a}_n\} \cap \icols}(q)
  }(\lbl{rewrite}{4}{rewrite-4})
\qquad
\inferrule
  {\relax
  }{
   q \cross 
   \begin{littbl}
     \begin{array}{|c|}
       \colhd{\scriptstyle a} \\
       \scriptstyle c \\
       \hline
     \end{array} 
   \end{littbl}
   \to \attach_{\col{a}:c}(q)
  }(\lbl{rewrite}{5}{rewrite-5})
\\[5mm]
\inferrule
  {\set
  }{
   \distinct(q) \to q
  }(\lbl{rewrite}{6}{rewrite-6})
\qquad
\inferrule
  {\relax
  }{
   \distinct(q)
   \to
   \distinct\left(\pi_{\var{cols}(q) \setminus (\const \setminus \icols)}(q)\right)
  }(\lbl{rewrite}{7}{rewrite-7})
\qquad
\inferrule
  {\neg\set \and \varoast \neq \distinct \and k \in \key \and k \subseteq \icols 
  }{
   \varoast(q) \to \distinct\left(\pi_{\icols}(\varoast(q))\right)
  }(\lbl{rewrite}{8}{rewrite-8})
\\[5mm]
\inferrule
  {\{\col{a}\} \in \key \and 
   q_1 \Rrightarrow q_2 \and
   q_2 \Rrightarrow q_1
  }{
   q_1 \Join_{\col{a} = \col{a}} q_2 \to q_1
  }(\lbl{rewrite}{9}{rewrite-9})
\quad\,\,
\inferrule
  {\{\col{a} = c, \col{b} = c\} \subseteq \const
  }{
   q_1 \Join_{\col{a} = \col{b}} q_2 
   \to
   q_1 \cross q_2
  }(\lbl{rewrite}{10}{rewrite-10})
\quad\,\,
\inferrule
  {\varoast \notin \{\distinct,\num\} \and 
   q_2 \not\Rrightarrow \varoast \and
   \{\col{a},\col{b}\} \subseteq \var{cols}(q_1) \cup \var{cols}(q_2)
  }{
   \varoast(q_1) \Join_{\col{a} = \col{b}} q_2
   \to
   \varoast(q_1 \Join_{\col{a} = \col{b}} q_2)
  }(\lbl{rewrite}{11}{rewrite-11})
\\[5mm]
\inferrule
  {\relax
  }{
   \rank_{\col{a}:\left<\col{b}\right>}(q)
   \to
   \pi_{\col{a}:\col{b},\var{cols}(q)}(q)
  }(\lbl{rewrite}{12}{rewrite-12})
\qquad
\inferrule
  {\relax
  }{
   \rank_{\col{a}:\left<\col{b}_1,\dots,\col{b}_n\right>}(q)
   \to
   \rank_{\col{a}:\left<\col{b}_1,\dots,\col{b}_n\right> 
          \setminus \const}(q)
  }(\lbl{rewrite}{13}{rewrite-13})
\qquad
\inferrule
  {\varoast \in \{\select_p,\distinct,\attach,\num\}
   \and
   \col{a} \notin \var{cols}(p)  
  }{
   \varoast(\rank_{\col{a}:\left<\col{b_1},\dots,\col{b_n}\right>}(q))
   \to
   \rank_{\col{a}:\left<\col{b}_1,\dots,\col{b}_n\right>}(\varoast(q)) 
  }(\lbl{rewrite}{14}{rewrite-14})
\\[5mm]
\inferrule
  {\varoast \in \{\Join_p,\cross\}
   \and
   \col{a} \notin \var{cols}(p)
  }{
   \rank_{\col{a}:\left<\col{b}_1,\dots,\col{b}_n\right>}(q_1) \varoast q_2
   \to
   \rank_{\col{a}:\left<\col{b}_1,\dots,\col{b}_n\right>}(q_1 \varoast q_2)
  }(\lbl{rewrite}{15}{rewrite-15})
\qquad
\inferrule
  {\relax
  }{
   \pi_{\col{a},\col{c}_1,\dots,\col{c}_m}\left(\rank_{\col{a}:\left<\col{b}_1,\dots,\col{b}_n\right>}(q)\right)
   \to
   \rank_{\col{a}:\left<\col{b}_1,\dots,\col{b}_n\right>}\left(\pi_{\col{b}_1,\dots,\col{b}_n,\col{c}_1,\dots,\col{c}_m}(q)\right)
  }(\lbl{rewrite}{16}{rewrite-16})
\\
\inferrule
  {\relax
  }{
   \rank_{\col{a}:\left<\dots,\col{b}_i,\dots\right>}\left(\rank_{\col{b}_i:\left<\col{c}_1,\dots,\col{c}_m\right>}(q)\right) 
   \to
   \rank_{\col{a}:\langle\dots,\col{b}_{i-1},
     \col{c}_1,\dots,\col{c}_m,\col{b}_{i+1},\dots\rangle}\left(
       \rank_{\col{b}_i:\left<\col{c}_1,\dots,\col{c}_m\right>}(q)\right)
  }(\lbl{rewrite}{17}{rewrite-17})
\end{array}
$$
\caption{Join graph isolation transformation (for a rule 
  $\var{lhs} \to \var{rhs}$, the properties \icols, \const, \set, and 
  \key{} denote the properties of $\var{lhs}$).}
\label{rewrite-rules}
\end{figure*}

A join of this type preserves the keys established by
$\num_{\col{inner}}$ and thus emits unique rows.  The introduction of a
new $\distinct$ instance at the top of the plan fragment thus does not alter
the plan semantics (Rule~\eqref{rewrite-8}, see
Figures~\ref{join-pushdown}\subref{join-pushdown-1} and
\subref{join-pushdown-2}).  This renders the original instance of
$\distinct$ obsolete as duplicate elimination now occurs upstream 
(Rule~\eqref{rewrite-6}, Fig.~\ref{join-pushdown}\subref{join-pushdown-3}).
The following stages push the join towards the plan base, leaving
a trail of plan sections that formerly occurred in the join input branches
(Rule~\eqref{rewrite-11}, Figures~\ref{join-pushdown}\subref{join-pushdown-3}
and \subref{join-pushdown-4}).   The condition $q_{2} \not\Rrightarrow
\varoast$ in the premise of Rule~\eqref{rewrite-11} prevents its further 
application in
the situation of Fig.~\ref{join-pushdown}\subref{join-pushdown-4}:
otherwise, the rewrite would introduce a cycle in the plan.  Instead,  
Rule~\eqref{rewrite-9} detects that the join has degenerated into a key
join over identical inputs and thus may be removed
(Fig.~\ref{join-pushdown}\subref{join-pushdown-5}).  Finally, this renders
the remaining instance of $\num_{\col{inner}}$ obsolete 
(column $\col{inner}$ not referenced, Rule~\eqref{rewrite-1}). 

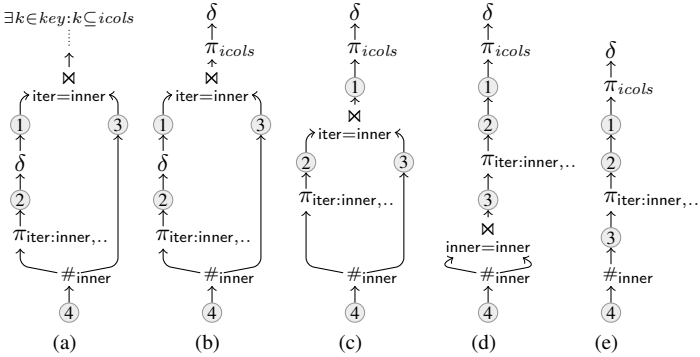
\begin{figure}[t]
  \small%
  \subfigure[]{%
  \label{join-pushdown-1}%
  \begin{tikzpicture}[x=6.5mm,y=5mm,inner sep=1pt,rounded corners=3pt,<-]
    \tikzstyle{plan}=[circle,inner sep=0.5pt,
                      draw=black!40,fill=black!7,font=\scriptsize]
    \draw
    (0,10)   node       (root)     {}
    (0,9)    node       (join)     {$\underset{\col{iter} = \col{inner}}{\Join}$}
    (-1,8)   node[plan] (plan1)    {1}
    (1,8)    node[plan] (plan3)    {3}
    (-1,7)   node       (distinct) {\distinct}
    (-1,6)   node[plan] (plan2)    {2}
    (-1,5)   node       (pi)       {$\pi_{\mathrlap{\col{iter}:\col{inner},\ndots}}$}
    (0,4)    node       (num)      {$\num_{\mathrlap{\col{inner}}}$}
    (0,3)    node[plan] (plan4)    {4}
             node[node distance=4mm,above of=root] (key)
                                   {$\mathclap{\scriptstyle \exists k \in \key: k \subseteq \icols}$}
    ;
    \draw (root) -- (join);
    \draw (join) -- +(-1,-0.25) -- (plan1);
    \draw (join) -- +( 1,-0.25) -- (plan3);
    \draw (plan1) -- (distinct);
    \draw (distinct) -- (plan2);
    \draw (plan2) -- (pi);
    \draw (pi) -- +(0,-0.75) -- (num);
    \draw (num) -- (plan4);
    \draw (plan3) -- +(0,-3.75) -- (num);
    \draw[densely dotted,-](key) -- (root);
  \end{tikzpicture}}
  \hfil
  \subfigure[]{%
  \label{join-pushdown-2}%
  \begin{tikzpicture}[x=6.5mm,y=5mm,inner sep=1pt,rounded corners=3pt,<-]
    \tikzstyle{plan}=[circle,inner sep=0.5pt,
                      draw=black!40,fill=black!7,font=\scriptsize]
    \draw         
    (0,11)   node       (ndistinct) {$\distinct$}
    (0,10)   node       (icols)     {$\pi_{\mathrlap{\icols}}$}
    (0,9)    node       (join)      {$\underset{\col{iter} = \col{inner}}{\Join}$}
    (-1,8)   node[plan] (plan1)     {1}
    (1,8)    node[plan] (plan3)     {3}
    (-1,7)   node       (distinct)  {\distinct}
    (-1,6)   node[plan] (plan2)     {2}
    (-1,5)   node       (pi)        {$\pi_{\mathrlap{\col{iter}:\col{inner},\ndots}}$}
    (0,4)    node       (num)       {$\num_{\mathrlap{\col{inner}}}$}
    (0,3)    node[plan] (plan4)     {4}
    ;
    \draw (ndistinct) -- (icols);
    \draw (icols) -- (join);
    \draw (join) -- +(-1,-0.25) -- (plan1);
    \draw (join) -- +( 1,-0.25) -- (plan3);
    \draw (plan1) -- (distinct);
    \draw (distinct) -- (plan2);
    \draw (plan2) -- (pi);
    \draw (pi) -- +(0,-0.75) -- (num);
    \draw (num) -- (plan4);
    \draw (plan3) -- +(0,-3.75) -- (num);
  \end{tikzpicture}}
  \hfil
  \subfigure[]{%
  \label{join-pushdown-3}%
  \begin{tikzpicture}[x=6.5mm,y=5mm,inner sep=1pt,rounded corners=3pt,<-]
    \tikzstyle{plan}=[circle,inner sep=0.5pt,
                      draw=black!40,fill=black!7,font=\scriptsize]
    \draw         
    (0,11)   node       (ndistinct) {$\distinct$}
    (0,10)   node       (icols)     {$\pi_{\mathrlap{\icols}}$}
    (0,8)    node       (join)      {$\underset{\col{iter} = \col{inner}}{\Join}$}
    (0,9)    node[plan] (plan1)     {1}
    (1,7)    node[plan] (plan3)     {3}
    (-1,7)   node[plan] (plan2)     {2}
    (-1,6)   node       (pi)        {$\pi_{\mathrlap{\col{iter}:\col{inner},\ndots}}$}
    (0,4)    node       (num)       {$\num_{\mathrlap{\col{inner}}}$}
    (0,3)    node[plan] (plan4)     {4}
    ;
    \draw (ndistinct) -- (icols);
    \draw (icols) -- (plan1);
    \draw (plan1) -- (join);
    \draw (join) -- +(-1,-0.25) -- (plan2);
    \draw (join) -- +( 1,-0.25) -- (plan3);
    \draw (plan2) -- (pi);
    \draw (pi) -- +(0,-1.75) -- (num);
    \draw (num) -- (plan4);
    \draw (plan3) -- +(0,-2.75) -- (num);
  \end{tikzpicture}%
  }
  \hfil
  \subfigure[]{%
  \label{join-pushdown-4}%
  \begin{tikzpicture}[x=6.5mm,y=5mm,inner sep=1pt,rounded corners=4pt,<-]
    \tikzstyle{plan}=[circle,inner sep=0.5pt,
                      draw=black!40,fill=black!7,font=\scriptsize]
    \draw         
    (0,11)   node       (ndistinct) {$\distinct$}
    (0,10)   node       (icols)     {$\pi_{\mathrlap{\icols}}$}
    (0,9)    node[plan] (plan1)     {1}
    (0,8)    node[plan] (plan2)     {2}
    (0,7)    node       (pi)        {$\pi_{\mathrlap{\col{iter}:\col{inner},\ndots}}$}
    (0,6)    node[plan] (plan3)     {3}
    (0,5)    node       (join)      {$\underset{\col{inner} = \col{inner}}{\Join}$}
    (0,4)    node       (num)       {$\num_{\mathrlap{\col{inner}}}$}
    (0,3)    node[plan] (plan4)     {4}   
    ;
    \draw (ndistinct) -- (icols);
    \draw (icols) -- (plan1);
    \draw (plan1) -- (plan2);
    \draw (plan2) -- (pi);
    \draw (pi) -- (plan3);
    \draw (plan3) -- (join);
    \draw (join) -- +(-1,-0.75) -- (num);
    \draw (join) -- +( 1,-0.75) -- (num);
    \draw (num) -- (plan4);
  \end{tikzpicture}%
  }
  \hfil
  \subfigure[]{%
  \label{join-pushdown-5}%
  \begin{tikzpicture}[x=6.5mm,y=5mm,inner sep=1pt,rounded corners=4pt,<-]
    \tikzstyle{plan}=[circle,inner sep=0.5pt,
                      draw=black!40,fill=black!7,font=\scriptsize]
    \draw         
    (0,10)   node       (ndistinct) {$\distinct$}
    (0,9)    node       (icols)     {$\pi_{\mathrlap{\icols}}$}
    (0,8)    node[plan] (plan1)     {1}
    (0,7)    node[plan] (plan2)     {2}
    (0,6)    node       (pi)        {$\pi_{\mathrlap{\col{iter}:\col{inner},\ndots}}$}
    (0,5)    node[plan] (plan3)     {3}
    (0,4)    node       (num)       {$\num_{\mathrlap{\col{inner}}}$}
    (0,3)    node[plan] (plan4)     {4}
    ;
    \draw (ndistinct) -- (icols);
    \draw (icols) -- (plan1);
    \draw (plan1) -- (plan2);
    \draw (plan2) -- (pi);
    \draw (pi) -- (plan3);
    \draw (plan3) -- (num);
    \draw (num) -- (plan4);
    \draw [draw=none,use as bounding box] (-1,3) rectangle (1,11);
  \end{tikzpicture}%
  }
  \caption{Moving duplicate elimination into the plan tail
    and join push-down (stages shown left to right).}
  \label{join-pushdown}
\end{figure}

\subsection{\XQuery{} in the Guise of \SQL{} SFW--Blocks}
\label{xquery-in-the-guise-of-sfw}

\begin{figure}[t]
  \centering\small  
\begin{tikzpicture}[scale=0.5,>=to,join=miter,y=0.28mm,x=0.45mm]
        \draw [densely dotted] (-20,280) -- (325,280);   \node at (291,300) {\emph{plan tail}};   \node at (285,255) {\emph{join graph}}; %
\node (ser5_) at (101,411) [draw,draw=none] {$\relax$};
  \node (ddo25_1) at (101,360) [draw,draw=none] {$\distinct$};
  \node (doc5_) at (143,4) [draw,draw=none] {$\doc$};
  \node (ddo25_2) at (101,306) [draw,draw=none] {$\pi_{\col{pre},\col{size},\col{level},\col{kind},\col{name},\col{value}}$};
  \node (child5_3) at (101,248) [draw,draw=none] {$\underset{                                                (\col{pre} < \col{pre}_1 \leqslant \col{pre} + \col{size}) \mathbin{\wedge}        (\col{level} + 1 = \col{level}_1)}{\Join}$};
  \node (child5_4) at (42,189) [draw,draw=none] {$\pi_{\renewcommand{\arraystretch}{0.1}                                       \begin{array}{@{}l@{}}                            											  \scriptstyle                                                            \col{pre}_1:\col{pre},\col{level}_1:\col{level}                       \end{array}}$};
  \node (child5_5) at (48,62) [draw,draw=none] {$\select_{\renewcommand{\arraystretch}{0.1}                                  \begin{array}{@{}l@{}}                   													  \scriptstyle                                                       (\col{kind} = \kind{ELEM}) \mathbin{\wedge}                              \\                                   													  \scriptstyle                                                       (\col{name} = \xquery{'bidder'})                                 \end{array}}$};
  \node (descendant5_3) at (160,189) [draw,draw=none] {$\underset{\col{pre}_{\circ} < \col{pre} \leqslant \col{pre}_{\circ} + \col{size}_{\circ}}{\Join}$};
  \node (descendant5_4) at (143,127) [draw,draw=none] {$\select_{\renewcommand{\arraystretch}{0.1}                                                \begin{array}{@{}l@{}}                            															   \scriptstyle                                                                     (\col{kind} = \kind{ELEM}) \mathbin{\wedge}                                      \\                                            															   \scriptstyle                                                                     (\col{name} = \xquery{'open\_auction'})                                        \end{array}}$};
  \node (descendant5_5) at (265,127) [draw,draw=none] {$\pi_{\renewcommand{\arraystretch}{0.1}                                                      \begin{array}{@{}l@{}}                                      														 \scriptstyle                                                                           \col{pre}_{\circ}:\col{pre},\col{size}_{\circ}:\col{size}                            \end{array}}$};
  \node (doc5_3) at (253,62) [draw,draw=none] {$\select_{\renewcommand{\arraystretch}{0.1}                                 \begin{array}{@{}l@{}}                    												  \scriptstyle                                                      (\col{kind} = \kind{DOC}) \wedge {}                               \\                                    												  \scriptstyle                                                      (\col{name} = \xquery{'auction.xml'})                          \end{array}}$};
  \draw [<-,o-] (ser5_) -- (ddo25_1);
  \draw [<-] (ddo25_1) -- (ddo25_2);
  \draw [<-] (ddo25_2) -- (child5_3);
  \draw [<-] (child5_3) -- (child5_4);
  \draw [<-] (child5_4) -- (child5_5);
  \draw [<-] (child5_5) -- (doc5_);
  \draw [<-] (child5_3) -- (descendant5_3.north);
  \draw [<-] (descendant5_3) -- (descendant5_4);
  \draw [<-] (descendant5_3) -- (descendant5_5);
  \draw [<-] (descendant5_4) -- (doc5_);
  \draw [<-] (descendant5_5) -- (doc5_3);
  \draw [<-] (doc5_3) -- (doc5_);
\end{tikzpicture}
  \newsavebox{\joingraphmark}
  \sbox{\joingraphmark}{%
    \tikz[baseline=-1mm,x=1ex] 
      \draw [densely dotted] (0,0) -- (3,0);}
  \caption{Final plan emitted for~\ref{Q1}. 
    The \usebox{\joingraphmark} separates the plan tail (above) from the isolated
    join bundle (three-fold self-join of table~\doc).}
  \label{plan-Q1-opt}
\end{figure}
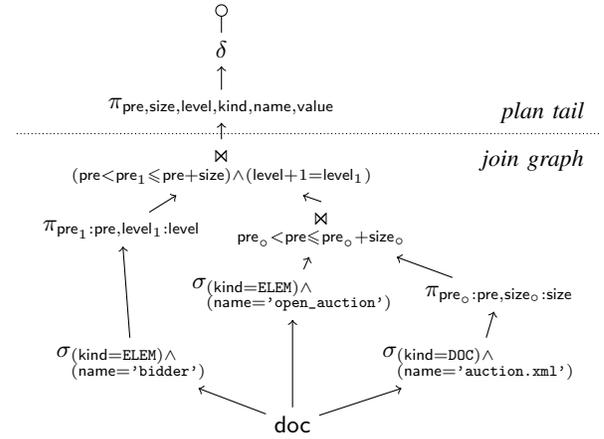

Fig.~\ref{plan-Q1-opt} depicts the isolation result for Query~\ref{Q1}
(original plan shown in Fig.~\ref{plan-Q1}). The new plan features a
bundle of operators in which---besides instances of $\pi$,
$\sigma$---the only remaining joins originate from applications of
compilation Rule~\ref{compile-Step}, implementing the semantics of
\XPath{} location steps. The joins consume rows from the \XML{} infoset
encoding table~$\doc$ which now is the only shared plan node in the DAG.
As desired, we can also identify the plan tail (in the case of~\ref{Q1},
no extra row ranking is required since the document order ranks of the
elements resulting from the \xquery{descendant::open\_auction} step---in
column $\col{pre}$ produced by the topmost $\pi$ operator---already
determine the overall order of the result.)

Quite unlike the initial plans emitted by the compositional compiler,
\XQuery{} join graph isolation derives plans that are truly
indistinguishable from the algebraic plans produced by a regular \SQL{}
translator. We thus let an off-the-shelf relational database back-end
\emph{autonomously} take over from here. It is now reasonable to expect the
system to excel at the evaluation of the considered \XQuery{} fragment
as it will face a familiar workload. (This is exactly what we observe in
Section~\ref{db2-udb-xquery-wizard-builtin}.)

Most importantly, the join graphs provide a \emph{complete} description
of the input query's true \XQuery{} semantics but \emph{do not
prescribe} a particular order of \XPath{} location step or predicate
evaluation. It is our intention to let the RDBMS decide on an
evaluation strategy, based on its very own cost model, the
availability of join algorithms, and supporting index structures. As a
consequence, it suffices to communicate the join graph in form of a
standard \SQL{} \SELECT-\DISTINCT-\FROM-\WHERE-\ORDERBY--block---\emph{i.e.},
in a de\-cla\-ra\-tive fashion barring any \XQuery{}-specific annotations or
similar clues. For Query~\ref{Q1}, we thus ship the \SQL{} query~\ref{Q1-SQL}
(Fig.~\ref{Q1-SQL-text}) for execution by the database back-end.

\begin{figure}[t]
  \centering\small
\begin{equation*}
\begin{BVerbatim}[baseline=c,commandchars=\\\{\},codes={\catcode`$=3}]
SELECT DISTINCT d2.*
  FROM $\doc$ AS d1, $\doc$ AS d2, $\doc$ AS d3
 WHERE d1.kind = DOC
   AND d1.name = 'auction.xml'
   AND d2.kind = ELEM
   AND d2.name = 'open_auction'
   AND d2 BETWEEN d1.pre\,+\,1 AND d1.pre\,+\,d1.size
   AND d3.kind = ELEM
   AND d3.name = 'bidder'
   AND d3 BETWEEN d2.pre\,+\,1 AND d2.pre\,+\,d2.size
   AND d2.level\,+\,1 = d3.level
 ORDER BY d2.pre
\end{BVerbatim}
\mskip10mu(\lbl{query}{\ensuremath{Q_{1}^{\sql{SQL}}}}{Q1-SQL})
\end{equation*}
\caption{\SQL{} encoding of \ref{Q1}'s join graph.}
\label{Q1-SQL-text}
\end{figure}

\begin{figure}[t]
  \centering\small
\begin{equation*}
\begin{BVerbatim}[baseline=c,commandchars=\\\{\},codes={\catcode`$=3}]
SELECT DISTINCT d12.*,
       d2.pre AS item1, d4.pre AS item2,
       d5.pre AS item3
  FROM $\doc$ AS d1, $\dots$, $\doc$ AS d12
 WHERE $\dots$
 ORDER BY d2.pre, d4.pre, d5.pre, d12.pre
\end{BVerbatim}
\mskip10mu(\lbl{query}{\ensuremath{Q_{2}^{\sql{SQL}}}}{Q2-SQL})
\end{equation*}
\caption{\SQL{} encoding of \ref{Q2} (focus on plan tail: order, duplicate removal).}
\label{Q2-SQL-text}
\end{figure}
\smallskip\noindent
\textbf{Plan tail.}
The interaction of \xquery{for} loop iteration and sequence
order of the final result becomes apparent in the plan tail of the
following query (traversing XMark data \cite{XMark} to return  
the names of those auction categories in which 
expensive items were sold at prices beyond 
{\fontencoding{T1}\fontfamily{rm}\textdollaroldstyle}$500$):
\begin{equation*}
\begin{BVerbatim}[baseline=c,commandchars=\\\{\},codes={\catcode`$=3}]
let \vardollar{}a := doc\,("auction.xml")
for \vardollar{}ca in \vardollar{}a//closed_auction[price > 500],
    \vardollar{}i  in \vardollar{}a//item,
    \vardollar{}c  in \vardollar{}a//category
where \vardollar{}ca/itemref/@item = \vardollar{}i/@id
  and \vardollar{}i/incategory/@category = \vardollar{}c/@id
return \vardollar{}c/name
\end{BVerbatim}
(\lbl{query}{\ensuremath{Q_{2}}}{Q2})
\end{equation*}
This example features \xquery{let} bindings and general value
com\-pa\-ri\-sons of atomized nodes, two extensions of the language
fragment of Fig.~\ref{grammar} that have been shown to readily fit
into the loop-lifting compilation approach \cite{XQueryOnSQLHosts}.
\XQuery{}~Core nor\-ma\-li\-za\-tion, compilation and subsequent
isolation yields the \SQL{} join graph query in Fig.~\ref{Q2-SQL-text}
which describes a 12-fold self-join over table~$\doc$. Note how the
\ORDERBY{} and \DISTINCT{} clauses---which represent the plan
tail---reflect the \XQuery{} sequence order and duplicate semantics:
\begin{compactenum}
\item[\emph{Order}] The nesting of the three \xquery{for} loops
  in~\ref{Q2} principally determines the order of the resulting node
  sequence: in~\ref{Q2}, row variables \sql{d2}, \sql{d4}, \sql{d5} range over
  \xquery{closed\_auction}, \xquery{item}, \xquery{category} element nodes,
  respectively.  The document order of the \xquery{name} elements 
  (bound to~\sql{d12}) is least relevant in this example and orders
  the nodes within each iteration of the innermost \xquery{for} loop.
\item[\emph{Duplicates}] The \XPath{} location step semantics requires
  duplicate node removal (row variable \sql{d12} appears in the
  \DISTINCT{} clause).  Duplicates are retained, however, across \xquery{for}
  loop iterations (keys \sql{d2.pre}, \sql{d4.pre}, and \sql{d5.pre} appear
  in the \DISTINCT{} clause).
\end{compactenum}

\section{In Laboratory with IBM\,DB2\,V9 (Experiments)}
\label{db2-udb-xquery-wizard-builtin}          

\newcommand{\ixinclude}{\smash{\ensuremath{\mskip-1mu\mid\mskip-1mu}}}

\definecolor{ixaccess} {rgb}{0.95,0.95,0.69}
\definecolor{ixscan}   {rgb}{0.95,0.95,0.69}
\definecolor{hsjoin}   {rgb}{0.80,0.69,0.66}
\definecolor{nljoin}   {rgb}{0.66,0.99,1}
\definecolor{sort}     {rgb}{1,0.56,0.56}
\definecolor{tbscan}   {rgb}{1,0.72,0.86}
\definecolor{return}   {rgb}{1,1,1}
\definecolor{tblaccess}{rgb}{1,0.72,0.86}

\newcommand{\ixaccess}[1]{%
  \begin{tikzpicture}[inner sep=3pt,outer sep=0pt]
    \node (n) [overlay] {\phantom{\ensuremath{\col{#1}}}};
    \draw [fill=ixaccess,overlay]
          (n.west)  -- (node cs:name=n,angle= 130)  --
          (n.north) -- (node cs:name=n,angle=  50)  --
          (n.east)  -- (node cs:name=n,angle= -50)  --
          (n.south) -- (node cs:name=n,angle=-130)  -- cycle;
    \node [overlay] {\ensuremath{\col{#1}}};
    \useasboundingbox [draw=none] (-3.5,-2) rectangle (3.5,2);
  \end{tikzpicture}
}
\newcommand{\ixscan}[2][ixscan]{%
  \begin{tikzpicture}[inner sep=2pt,outer sep=1pt]
    \node (n) {\phantom{\ensuremath{\col{#2}}}};
    \draw [fill=#1]
          (n.west)  -- (node cs:name=n,angle= 167)  -- 
                       (node cs:name=n,angle= 153)  -- 
          (n.north) -- (node cs:name=n,angle=  27)  --
                       (node cs:name=n,angle=  13)  --
          (n.east)  -- (node cs:name=n,angle= -13)  --
                       (node cs:name=n,angle= -27)  --
          (n.south) -- (node cs:name=n,angle=-153)  --
                       (node cs:name=n,angle=-167)  -- cycle;
    \node {\ensuremath{\col{#2}}};
  \end{tikzpicture}
}
\newcommand{\hsjoin}[1]{\ixscan[hsjoin]{#1}}
\newcommand{\nljoin}[1]{\ixscan[nljoin]{#1}}
\newcommand{\sort}[1]{\ixscan[sort]{#1}}
\newcommand{\tbscan}[1]{\ixscan[tbscan]{#1}}
\newcommand{\return}[1]{\ixscan[return]{#1}}
\newcommand{\tblaccess}[1]{%
  \begin{tikzpicture}[inner sep=2pt,outer sep=1pt]
    \node [draw,fill=tblaccess] {\ensuremath{\col{#1}}};
  \end{tikzpicture}
}

The \SQL{} language subset used to describe the \XQuery{} join
graphs---flat self-join chains, simple ordering criteria, and no
grouping or aggregation---is sufficiently simple to let any
\SQL{}-capable RDBMS assume the role of a back-end for \XQuery{}
evaluation. We do \emph{not} rely on \SQL{}/\XML{} functionality, in
particular. In what follows, we will observe how IBM\,DB2\,UDB\,V9 acts
as a runtime for the join-graph-isolating compiler. In this context,
DB2\,V9 appears to be especially interesting, because the system
\begin{compactenum}[(1)]
\item has the ability to \emph{autonomously} adapt the design of 
  its physical layer, indexes in particular, in response to a given 
  workload (this will help to assess whether an RDBMS can
  indeed cope with \XQuery{} specifics, like the pervasive sequence
  and document order notions), and
\item features the \emph{built-in} \XQuery{} processor \pureXML{} which
  implements a ``native'' \XML{} document storage and specific primitives
  for \XPath{} evaluation,
  but nevertheless relies on the very same data\-base kernel infrastructure.
  This will provide an insightful point of reference for the performance
  assessment of Section~\ref{experiments}.
\end{compactenum}

\smallskip\noindent
\textbf{Autonomous index design.} The workload produced by the
join-graph-isolating compiler is completely regular: as long as the
input expressions adhere to the \XQuery{} dialect of
Fig.~\ref{grammar}, \emph{all} emitted queries will, for example,
evaluate predicates against ranges with endpoints $\col{pre}$,
$\col{pre} + \col{size}$ and \emph{always} use column $\col{pre}$ as
ordering criterion which finds perfect support in a clustered B-tree on
the primary key column $\col{pre}$.  (Here, column $\col{size}$ is exclusively
used as a summand in $\col{pre} + \col{size}$---we thus replaced column $\col{size}$
with a computed column that contains the sum.)

Due to this high predictability, we expected the DB2 automatic design
advisor, \sql{db2advis} \cite{DB2V9}, to be able to suggest a
reasonable, tailored set of vanilla B-tree indexes to support the
typical \XQuery{} join graph workload. 

To provide the RDBMS with complete information about the expected
incoming queries, we instructed the compiler to make the semantics of
the serialization point $\ser$ explicit---this adds one extra
\xquery{descendant-or-self::node()} step to any Query~$Q$, originating
in its result node sequence: 
$$ 
\xquery{for \vardollar{}x in $Q$ return
\vardollar{}x/descendant-or-self::node()}\,. 
$$ 
This produces all \XML{} nodes required to fully serialize the result
(surfacing as the additional topmost self-join in the join plans of
Figures~\ref{join-plan-Q1} and~\ref{join-plan-Q2}).

\begin{table}
  \tikzstyle{lolepop}=[y=1.0mm,x=0.75mm,font=\scriptsize\sffamily,baseline]
  \centering\small            
  \caption{B-tree indexes proposed by \sql{db2advis}.}
  \label{index-proposal}
  \begin{tabular}{@{}lp{0.55\linewidth}@{}}
    \toprule
    \textbf{Index key columns} & \textbf{Index deployment} \\
    \cmidrule(r){1-1}
    \cmidrule(l){2-2}
    \tikz[lolepop] \node {\ixaccess{nkspl}};~  
    \tikz[lolepop] \node {\ixaccess{nlkps}};~  
    \tikz[lolepop] \node {\ixaccess{nksp}};~  
    \tikz[lolepop] \node {\ixaccess{nlkp}}; 
    &  
    \XPath{} node test and axis step, \newline
    access document node (\xquery{doc($\cdot$)})
    \\
    \tikz[lolepop] \node {\ixaccess{vnlkp}};~    
    \tikz[lolepop] \node {\ixaccess{nlkpv}};~   
    \tikz[lolepop] \node {\ixaccess{nkdlp}};
    &
    Atomization, value comparison with \newline
    subsequent/preceding XPath{} step
    \\
    ~\tikz[lolepop] \node {\ixaccess{p\ixinclude{}nvkls}};~ 
    & 
    Serialization support (supplies \XML{} \newline infoset 
    in document order)    
    \\
    \bottomrule 
    \multicolumn{2}{@{}l@{}}{%
    $\col{p}${:}$\col{pre}$, $\col{s}${:}$\col{pre}+\col{size}$,
    $\col{l}${:}$\col{level}$, $\col{k}${:}$\col{kind}$, 
    $\col{n}${:}$\col{name}$, $\col{v}${:}$\col{value}$,
    $\col{d}${:}$\col{data}$}
  \end{tabular}
\end{table}

For~\ref{Q2} as an representative of the prototypical expected query
workload, the DB2 design advisor suggests the B-tree index set of
Table~\ref{index-proposal}.  Configured to exploit an unlimited
time and index space budget, \sql{db2advis} proposes indexes 
that add up to a total size of 300\,MB for a 110\,MB instance of the
XMark \xquery{auction.xml} document.  Due to the regularity of
the emitted \SQL{} code, the utility of the proposed indexes will be 
high for any workload that exhibits a significant fraction of \XQuery{}
join graphs.

\smallskip\noindent
\textbf{Partitioned B-tree index support for \XQuery{}.}
The majority of the index keys proposed in 
Table~\ref{index-proposal} are prefixed with \emph{low cardinality}
column(s), \emph{e.g.}, $\col{n}$, $\col{nk}$, or $\col{nlk}$.  An
XMark \XML{} instance features 77~distinct element tag
and attribute names, regardless of the document size. Similar
observations apply to the 7~\XML{} node kinds (column~$\col{kind}$)
and the typical \XML{} document height ($0 \leqslant \col{level} \leqslant 14$ for
the XMark instances). A B-tree that
is primarily organized by such a low cardinality column will, in 
consequence, \emph{partition} the \XML{} infoset encoding 
into few disjoint node sets.  Note how, in a sense, 
a $\col{name}$-prefixed index key
leads to a B-tree-based implementation of the \emph{element tag 
streams}, the principal data access path used in the so-called
\emph{twig join} algorithms \cite{HolisticTwigJoin,Twig2Stack}.  Partitioned B-trees
enjoy a number of desirable properties \cite{PartitionedBtree}.  
In a $\col{name}$-partitioned B-tree with key prefix compression, 
for example, each partition will contain its element name exactly once
(zero redundancy tag name storage).

The design advisor further suggests an index with key $\col{vnlkp}$
whose $\col{value}$ column prefix supports atomization and the general
value comparisons between (attribute) nodes featured in~\ref{Q2}. A
B-tree of this type bears some close resemblance with the
\XPath{}-specific indexes (\sql{CREATE INDEX} $\dots$ \sql{GENERATE KEY
USING XMLPATTERN} $\dots$ \sql{AS SQL VARCHAR($n$)}) employed by
\pureXML{} (Section~\ref{experiments}).

The proposal of the unique clustered index with key
$\col{p\ixinclude{}nvkls}$ is owed to the
\xquery{descendant-or-self::node()} step, introduced to enable \XML{}
serialization. A forward scan of this index provides access to the full
\XML{} infoset for any subtree encoded in table~$\doc$, in document
order.  Since \sql{db2advis} has observed that such scans will be
the principal (here: only) use of this index, it proposes 
$\col{pre}$ as the only key column---all other columns ($\col{name}$
through $\col{pre} + \col{size}$) are specified
in DB2's \sql{INCLUDE($\cdot$)} index creation clause \cite{DB2V9}
and thus merely occupy space on the index leaf pages.

\subsection{\XPath{} Continuations}
\label{xpath-continuations}
How exactly does DB2\,V9's query optimizer deploy the indexes proposed
by its design advisor companion? An answer to this question can be found
through an analysis of the plan trees generated by the optimizer. We
have, in fact, observed a few not immediately obvious ``tricks'' that
have found their way into the execution plans. Most of these
observations are closely related to query evaluation techniques that
have originally been described as \XPath{}--specific
\cite{HolisticTwigJoin,QueryOptimizationforXML, LookingForward}, outside
the relational domain. Since we have transferred all responsibility for
the \XQuery{} runtime aspects to the RDBMS, we think this is quite
interesting.

The optimized DB2 execution plan found for Query~\ref{Q1} 
of Section~\ref{impact-of-compositionality} is shown in
Fig.~\ref{join-plan-Q1}.  We are reproducing these execution plans
in a form closely resembling the output of DB2's visual explain facility.  
Nodes in these plans represent operators of DB2's variant of physical
algebra---all operators relevant for the present discussion 
are introduced in Table~\ref{lolepops}.

\begin{figure}
\centering\scriptsize\sffamily
\begin{tikzpicture}[node distance=5mm,outer sep=0pt,inner sep=0pt,
       scale=0.4,>=latex,join=bevel,y=0.975mm,x=0.75mm]
  \node (node01t) at ( 65,-10) {\tblaccess{\doc}};
  \node (node02t) at (170,-10) {\tblaccess{\doc}};
  \node (node03t) at (175, 10) {\tblaccess{\doc}};
  \node (node04t) at (180, 30) {\tblaccess{\doc}};

  \node (node01)  at ( 65, 20) {\ixscan{IXSCAN}};
  \node (node02)  at (110, 20) {\ixscan{IXSCAN}};
  \node (node03)  at (115, 40) {\ixscan{IXSCAN}};
  \node (node04)  at (120, 60) {\ixscan{IXSCAN}};

  \node (node02x) at ( 70, 40) {\nljoin{NLJOIN}};
  \node (node03x) at ( 75, 60) {\nljoin{NLJOIN}};
  \node at (60,68) {$\ltimes$};
  \node (node04x) at ( 80, 80) {\nljoin{NLJOIN}};

  \node (node01i) at ( 65,  5) {\ixaccess{nksp}};
  \node (node02i) at (140,  5) {\ixaccess{nkspl}};
  \node (node03i) at (145, 25) {\ixaccess{nlkp}};
  \node (node04i) at (150, 45) {\ixaccess{p\ixinclude{}nvkls}};

  \node (node05)  at ( 80, 95) {\sort{SORT}};
  \node (node06)  at ( 80,110) {\tbscan{TBSCAN}};
  \node (node07)  at ( 80,125) {\return{RETURN}};

  \draw [->] (node01)  -- (node02x);
  \draw [->] (node02)  -- (node02x);
  \draw [->] (node02x) -- (node03x);
  \draw [->] (node03)  -- (node03x);
  \draw [->] (node03x) -- (node04x);
  \draw [->] (node04)  -- (node04x);
  
  \draw [->] (node04x) -- (node05);
  \draw [->] (node05)  -- (node06);
  \draw [->] (node06)  -- (node07);

  \draw [->] (node01t) -- (node01i);
  \draw [->] (node02t) -- (node02i);
  \draw [->] (node03t) -- (node03i);
  \draw [->] (node04t) -- (node04i);

  \draw [->] (node01i) -- (node01);
  \draw [->] (node02i) -- (node02);
  \draw [->] (node03i) -- (node03);
  \draw [->] (node04i) -- (node04);
  
  \node [below of=node01t,anchor=west,xshift=-3mm]
  {$\xquery{doc("auction.xml")}\anticip{\xquery{/descendant}}_{1}$};
  \node [right of=node02t,anchor=west]
  {$\anticip{\phantom{X}}_{1}\xquery{\!::open\_auction}\left\{
    \begin{array}{@{}l@{}}
      \anticip{\xquery{/d-o-s}}_{2} \\
      \anticip{\xquery{/child}}_{3}
    \end{array}\right.$};
  \node [right of=node03t,anchor=west]
  {$\anticip{\phantom{X}}_{3}\xquery{\!::bidder}$};
  \node [right of=node04t,anchor=west,yshift=-1mm]
  {$\begin{array}{@{}l}
     \anticip{\phantom{X}}_{2}\xquery{::node()} \\
     \textnormal{(serialization)}
     \end{array}
   $};
\end{tikzpicture}
\normalfont         
\caption{DB2\,V9 execution plan for~\ref{Q1} with continuation annotations
  (\xquery{d-o-s} abbreviates \xquery{descendant-or-self}).}
\label{join-plan-Q1}
\end{figure}

\begin{table}
  \tikzstyle{lolepop}=[y=1.0mm,x=0.75mm,font=\scriptsize\sffamily,baseline]
  {\centering\small                              
  \caption{Relevant IBM\,DB2 plan operators.}
  \label{lolepops}
  \begin{tabular}{@{}cp{0.255\linewidth}@{}ccp{0.27\linewidth}@{}}
    \toprule
    \textbf{Operator} & \textbf{Semantics} & &
    \textbf{Operator} & \textbf{Semantics} \\
    \cmidrule{1-2}
    \cmidrule(l){4-5}
    \tikz[lolepop] \node {\return{RETURN}}; & 
    Result row \newline delivery & &
    \tikz[lolepop] \node {\sort{SORT}}; & 
    Sort rows (+ dup. \newline row elimination) 
    \\
    \tikz[lolepop] \node {\nljoin{NLJOIN}}; & 
    Nested-loop join \newline (left leg: outer) & &
    \tikz[lolepop] \node {\hsjoin{HSJOIN}}; & 
    Hash join \newline (left leg: probe)
    \\
    \tikz[lolepop] \node {\ixscan{IXSCAN}}; & 
    B-tree scan & &
    \tikz[lolepop] \node {\tbscan{TBSCAN}}; & 
    Temporary table \newline scan 
    \\
    \tikz[lolepop] \node {\ixaccess{nlkps}}; & 
    Index access & &
    \tikz[lolepop] \node {\tblaccess{\doc}}; & 
    \XML{} infoset \newline table access
    \\
    \bottomrule 
  \end{tabular}}
\end{table}

\smallskip\noindent
\textbf{Path stitching and branching.} 
Consider the B-tree index with key $\col{nksp}$. Due to its $\col{nk}$
prefix, this index primarily provides support for \XPath{} name and kind
tests. In the execution plan for~\ref{Q1}, the index is used to access
the requested document node ($\col{name} = \xquery{'auction.xml'} \wedge
\col{kind} = \xquery{DOC}$). Additionally, however, the index delivers the
infoset properties $\col{sp}$  and thus
provides all necessary information to step along the \XPath{} 
\xquery{descendant} axis---namely the interval 
$(\col{pre}, \col{pre} + \col{size}]$, see Fig.~\ref{kindtnamet}---from those nodes
that have been found during index lookup.  In
the following, we will denote the result of such index lookups as 
$$
  \xquery{doc("auction.xml")\anticip{/descendant}}_1
$$
or, generally $n\,\anticip{\xquery{/}$\alpha$}_{i}$ (read: perform the specified node test $n$, then prepare a subsequent
step along axis $\alpha$). In the execution plan of 
Fig.~\ref{join-plan-Q1}, the bottom index nested-loop join continues
this ``half-cooked'' step: a lookup in the index $\col{nkspl}$
retrieves columns $\col{nkp}$ to 
\begin{inparaenum}[(1)]
\item perform the due 
  name and kind test 
  ($\col{name} = \xquery{'open\_auction'} \wedge \col{kind} = \xquery{ELEM}$) and
\item complete the structural \xquery{descendant} axis traversal
  (check $\col{p}$ for containment in the $(\col{pre}, \col{pre} + \col{size}]$
  interval obtained in the first half of the step).                 
\end{inparaenum}
In the annotated plans, we write 
$\anticip{\phantom{C}}_{1}\xquery{::open\_auction}$ (read: resume axis
step and perform specified name and kind test).  Stitched at the
matching continuation points (here: those with subscript $1$), we
obtain the complete \XPath{} location step again:
$$
  \xquery{doc("auction.xml")\anticip{/descendant}}_1\!\xquery{::open\_auction}\enskip.
$$
The lookup in index $\col{nkspl}$ further provides the necessary
infoset properties to prepare the now current continuations 
$\xquery{\anticip{/descendant-or-self}}_{2}$ (columns $\col{sp}$) as well
as $\xquery{\anticip{/child}}_{3}$ (columns $\col{spl}$).  Such
continuations with multiple resumption points are the equivalent of
the branching nodes discussed in the context of \emph{holistic twig joins}
\cite{HolisticTwigJoin}.

Given the tailored B-tree index set in Table~\ref{index-proposal}, the
DB2 query optimizer consistently manages to select the index access path
that provides just the required \XML{} infoset properties. Resuming
the \xquery{child} continuation at
$\anticip{\phantom{C}}_{3}\xquery{::bidder}$ requires columns $\col{nk}$
to perform the name and kind test plus columns $\col{pl}$ to complete 
the evaluation of the range predicate that implements the
\xquery{child} step  (again, see Fig.~\ref{kindtnamet},
columns $\col{pre}_{\circ}$, $\col{size}_{\circ}$, $\col{level}_{\circ}$
are provided by the $\xquery{\anticip{/child}}_{3}$ continuation).
The optimizer thus selected the index with key $\col{nlkp}$.  Finally,
as anticipated, the plan scans index $\col{p\ixinclude{}nvkls}$ 
to traverse all nodes in the subtrees below the query's \XML{} result nodes 
(resuming from $\xquery{\anticip{/descendant-or-self}}_{2}$).

\smallskip\noindent
\textbf{\XPath{} step reordering and axis reversal.}
Two further phenomena in the execution plans can be explained in terms
of the \XPath{} continuation notion.  For~\ref{Q1}, the order of the 
\XPath{} location steps specified in the input query did coincide with
the join order in the execution plan (access document node of \XML{}
instance \xquery{auction.xml}, then perform a \xquery{descendant::open\_auction}
step, finally evaluate \xquery{child::bidder}).  The B-tree index entries
provide sufficient context information, however, to allow for arbitrary
path processing orders---right-to-left strategies are conceivable as
are strategies that start in the middle of a step sequence and then work
their way towards the path's endpoints.  The latter
can be witnessed in the execution plan of Query~\ref{Q2} 
(Fig.~\ref{join-plan-Q2}).

\begin{figure}[t]
\centering\scriptsize\sffamily
\begin{tikzpicture}[node distance=5mm,outer sep=0pt,inner sep=0pt,
       scale=0.4,>=latex,join=bevel,y=0.975mm,x=0.75mm]
  \node (node01t) at (175, 10) {\tblaccess{\doc}};
  \node (node02t) at ( 65,-10) {\tblaccess{\doc}};
  \node (node03t) at (170,-10) {\tblaccess{\doc}};
  \node (node05t) at (180, 30) {\tblaccess{\doc}};
  \node (node06t) at (185, 50) {\tblaccess{\doc}};
  \node (node07t) at (190, 70) {\tblaccess{\doc}};
  \node (node08t) at (195, 90) {\tblaccess{\doc}};
  \node (node09t) at (200,110) {\tblaccess{\doc}};
  \node (node10t) at (205,130) {\tblaccess{\doc}};
  \node (node11t) at (210,150) {\tblaccess{\doc}};
  \node (node12t) at (215,170) {\tblaccess{\doc}};
  \node (node13t) at (220,190) {\tblaccess{\doc}};
  \node (node14t) at (225,210) {\tblaccess{\doc}};
  \node (node01)  at (115, 40) {\ixscan{IXSCAN}};
  \node (node02)  at ( 65, 20) {\ixscan{IXSCAN}};
  \node (node03)  at (110, 20) {\ixscan{IXSCAN}};
  \node (node05)  at (120, 60) {\ixscan{IXSCAN}};
  \node (node06)  at (125, 80) {\ixscan{IXSCAN}};
  \node (node07)  at (130,100) {\ixscan{IXSCAN}};
  \node (node08)  at (135,120) {\ixscan{IXSCAN}};
  \node (node09)  at (140,140) {\ixscan{IXSCAN}};
  \node (node10)  at (145,160) {\ixscan{IXSCAN}};
  \node (node11)  at (150,180) {\ixscan{IXSCAN}};
  \node (node12)  at (155,200) {\ixscan{IXSCAN}};
  \node (node13)  at (160,220) {\ixscan{IXSCAN}};
  \node (node14)  at (165,240) {\ixscan{IXSCAN}};
  \node (node01x) at ( 75, 60) {\nljoin{NLJOIN}};
  \node (node03x) at ( 70, 40) {\hsjoin{HSJOIN}};
  \node at (55, 48) {$\ltimes$};
  \node (node05x) at ( 80, 80) {\nljoin{NLJOIN}};
  \node (node06x) at ( 85,100) {\nljoin{NLJOIN}};
  \node (node07x) at ( 90,120) {\nljoin{NLJOIN}};
  \node (node08x) at ( 95,140) {\nljoin{NLJOIN}};
  \node (node09x) at (100,160) {\nljoin{NLJOIN}};
  \node (node10x) at (105,180) {\nljoin{NLJOIN}};
  \node (node11x) at (110,200) {\nljoin{NLJOIN}};
  \node (node12x) at (115,220) {\nljoin{NLJOIN}};
  \node (node13x) at (120,240) {\nljoin{NLJOIN}};
  \node (node14x) at (125,260) {\nljoin{NLJOIN}};
  \node (node01i) at (145, 25) {\ixaccess{nksp}};
  \node (node02i) at ( 65,  5) {\ixaccess{nkspl}};
  \node (node03i) at (140,  5) {\ixaccess{nkdlp}};
  \node (node05i) at (150, 45) {\ixaccess{nlkps}};
  \node (node06i) at (155, 65) {\ixaccess{nlkpv}};
  \node (node07i) at (160, 85) {\ixaccess{vnlkp}};
  \node (node08i) at (165,105) {\ixaccess{nkspl}};
  \node (node09i) at (170,125) {\ixaccess{nlkps}};
  \node (node10i) at (175,145) {\ixaccess{nlkpv}};
  \node (node11i) at (180,165) {\ixaccess{vnlkp}};
  \node (node12i) at (185,185) {\ixaccess{nkspl}};
  \node (node13i) at (190,205) {\ixaccess{nlkps}};
  \node (node14i) at (195,225) {\ixaccess{p\ixinclude{}nvkls}};
  \node (node15)  at (125,275) {\sort{SORT}};
  \node (node16)  at (125,290) {\tbscan{TBSCAN}};
  \node (node17)  at (125,305) {\return{RETURN}};
  \draw [->] (node02)  -- (node03x);
  \draw [->] (node03)  -- (node03x);
  \draw [->] (node03x) -- (node01x);
  \draw [->] (node01)  -- (node01x);
  \draw [->] (node01x) -- (node05x);
  \draw [->] (node05)  -- (node05x);
  \draw [->] (node05x) -- (node06x);
  \draw [->] (node06)  -- (node06x);
  \draw [->] (node06x) -- (node07x);
  \draw [->] (node07)  -- (node07x);
  \draw [->] (node07x) -- (node08x);
  \draw [->] (node08)  -- (node08x);
  \draw [->] (node08x) -- (node09x);
  \draw [->] (node09)  -- (node09x);
  \draw [->] (node09x) -- (node10x);
  \draw [->] (node10)  -- (node10x);
  \draw [->] (node10x) -- (node11x);
  \draw [->] (node11)  -- (node11x);
  \draw [->] (node11x) -- (node12x);
  \draw [->] (node12)  -- (node12x);
  \draw [->] (node12x) -- (node13x);
  \draw [->] (node13)  -- (node13x);
  \draw [->] (node13x) -- (node14x);
  \draw [->] (node14)  -- (node14x);
  \draw [->] (node14x) -- (node15);
  \draw [->] (node15)  -- (node16);
  \draw [->] (node16)  -- (node17);
  \draw [->] (node01t) -- (node01i);
  \draw [->] (node02t) -- (node02i);
  \draw [->] (node03t) -- (node03i);
  \draw [->] (node05t) -- (node05i);
  \draw [->] (node06t) -- (node06i);
  \draw [->] (node07t) -- (node07i);
  \draw [->] (node08t) -- (node08i);
  \draw [->] (node09t) -- (node09i);
  \draw [->] (node10t) -- (node10i);
  \draw [->] (node11t) -- (node11i);
  \draw [->] (node12t) -- (node12i);
  \draw [->] (node13t) -- (node13i);
  \draw [->] (node14t) -- (node14i);
  \draw [->] (node01i) -- (node01);
  \draw [->] (node02i) -- (node02);
  \draw [->] (node03i) -- (node03);
  \draw [->] (node05i) -- (node05);
  \draw [->] (node06i) -- (node06);
  \draw [->] (node07i) -- (node07);
  \draw [->] (node08i) -- (node08);
  \draw [->] (node09i) -- (node09);
  \draw [->] (node10i) -- (node10);
  \draw [->] (node11i) -- (node11);
  \draw [->] (node12i) -- (node12);
  \draw [->] (node13i) -- (node13);
  \draw [->] (node14i) -- (node14);
  \node [right of=node01t,anchor=west] 
  {$\xquery{doc("auction.xml")}\anticip{\xquery{/descendant}}_{1}$};
  \node [below of=node02t,anchor=west,xshift=-3mm]
  {$\anticip{\phantom{X}}_{1}\xquery{\!::closed\_auction}\anticip{\xquery{/child}}_{2}$};
  \node [right of=node03t,anchor=west]
  {$\anticip{\phantom{X}}_{2}\xquery{\!::price[data(.)\,>\,500]}$};
  \node [right of=node05t,anchor=west]
  {$\anticip{\phantom{X}}_{2}\xquery{\!::itemref}\anticip{\xquery{/attribute}}_{3}$};
  \node [right of=node06t,anchor=west]
  {$\anticip{\phantom{X}}_{3}\xquery{\!::item}\anticip{\xquery{/data(.)}}_{4}$};
  \node [right of=node07t,anchor=west]
  {$\anticip{\phantom{X}}_{5}\xquery{\!::id/data(.)}\,\xquery{=}\,\anticip{\phantom{X}}_{4}$};
  \node [right of=node08t,anchor=west]
  {$\anticip{\phantom{X}}_{1}\xquery{\!::item}\left\{
    \begin{array}{@{}l@{}}
      \anticip{\xquery{/attribute}}_{5} \\
      \anticip{\xquery{/child}}_{6}
    \end{array}\right.
   $};
   \node [right of=node09t,anchor=west]
   {$\anticip{\phantom{X}}_{6}\xquery{\!::incategory}\anticip{\xquery{/attribute}}_{7}$};
   \node [right of=node10t,anchor=west]
   {$\anticip{\phantom{X}}_{7}\xquery{\!::category}\anticip{\xquery{/data(.)}}_{8}$};
   \node [right of=node11t,anchor=west]
   {$\anticip{\phantom{X}}_{9}\xquery{\!::id/data(.)}\,\xquery{=}\,\anticip{\phantom{X}}_{8}$};
   \node [right of=node12t,anchor=west]
   {$\anticip{\phantom{X}}_{1}\xquery{\!::category}\left\{
    \begin{array}{@{}l@{}}
      \anticip{\xquery{/attribute}}_{9} \\
      \anticip{\xquery{/child}}_{10}
    \end{array}\right.
    $};
   \node [right of=node13t,anchor=west]
   {$\anticip{\phantom{X}}_{10}\xquery{\!::name}\anticip{\xquery{/d-o-s}}_{11}$};
   \node [right of=node14t,anchor=west]
   {$\begin{array}{@{}l}
     \anticip{\phantom{X}}_{11}\xquery{::node()} \\
     \textnormal{(serialization)}
     \end{array}
    $};
\end{tikzpicture}
\normalfont
\caption{DB2\,V9 execution plan for~\ref{Q2}. }
\label{join-plan-Q2}
\end{figure}

The plan's very first index scan over $\col{nkspl}$ evaluates the name 
and kind test for elements with tag \xquery{closed\_auction} 
\emph{before} the continuation for
resumption point $\anticip{\phantom{C}}_{1}$ has provided any
context node information.  The index key columns $\col{spl}$ are 
used to prepare the continuation $\anticip{\xquery{/child}}_{2}$ which
is then immediately resumed by the node test for element 
\xquery{price}.  Index $\col{nkdlp}$ is deployed to implement this node
test as well as node atomization and subsequent value comparison 
(the index key contains column $\col{data}$ of table~$\doc$).  At this
point, the plan has evaluated the path fragment
$$
  \anticip{\phantom{C}}_{1}\!\xquery{::closed\_auction}%
  \anticip{\xquery{/child}}_{2}\!\xquery{::price[data(.)\,>\,500]}
$$
which is still context-less. The due context is only
provided by the subsequent $\col{NLJOIN}$--$\col{IXSCAN}$ pair which
verifies that the \xquery{closed\_auction} elements found so far
indeed are \xquery{descendant}s of \xquery{auction.xml}'s document node.  
Observe that in this specific evaluation order of the location
steps, the \xquery{closed\_auction} nodes now assume the context node
role: the plan effectively determines the \xquery{closed\_auction}
elements that have the document node of \xquery{auction.xml} in their
\xquery{ancestor} axis.  The reversal of axes---in this case, trading
\xquery{descendant} for \xquery{ancestor}---is based on the dualities
$\col{pre}\leftrightarrow\col{pre}_{\circ}$, 
$\col{size}\leftrightarrow\col{size}_{\circ}$
in the predicates $\var{axis}(\xquery{ancestor})$ and 
$\var{axis}(\xquery{descendant})$, defined in Fig.~\ref{kindtnamet}.
This observation applies to all other pairs of dual axes
\cite{LookingForward} (\emph{e.g.}, \xquery{parent}/\xquery{child},
\xquery{following}/\xquery{preceding}) and, due to the attribute
encoding used in table~$\doc$, also to the attribute/owner relationship
between element nodes and the attributes they own.

\smallskip\noindent
The query optimizer decides on the reordering of paths and the
associated reversal of \XPath{} axes based on its ``classical''
selectivity notion and the availability of eligible access paths: 
for a B-tree with $\col{name}$-prefixed keys, the
RDBMS's data distribution statistics capture tag name distribution while
$\col{value}$-prefixed keys lead to statistics about the distribution
of the (untyped) element and attribute values. 

In the case of~\ref{Q2}, this enabled the optimizer to decide that the
access path $\col{nkdlp}$, directly leading to \xquery{price} nodes
(key prefix $\col{nk}$) with a typed decimal value of greater than 500 (key
column $\col{d}$), is highly selective (only 9,750 of the 4.7~million
nodes in the 110\,MB XMark \XML{} instance are \xquery{price} elements
and only a fraction of these has a typed value in the required range).

An analogous observation about the distribution of untyped string values in 
the $\col{value}$ column---the key prefix in the $\col{vnlkp}$ B-tree---has
led the optimizer to evaluate the general attribute value comparison
$$    
  \begin{array}{@{}c@{}}
    \anticip{\phantom{C}}_{5}\!\xquery{::id/data(.)\,=\,} \\
    \xquery{doc("auction.xml")/$\cdots$/attribute::item/data(.)}
  \end{array}    
$$
before the ``hole'' $\anticip{\phantom{C}}_{5}$ has been filled. The
elements owning the \xquery{@id} attributes are resolved subsequently,
effectively reversing the \xquery{attribute} axis. (This constellation
repeats for the second attribute value comparison in~\ref{Q2},
resumption point $\anticip{\phantom{C}}_{9}$.)

\subsection{Pure \SQL{} \emph{vs.} pure\XML{\large\texttrademark}}
\label{experiments}     

With DB2\,Version~9, IBM released the built-in \pureXML{} \XQuery{}
processor. This opens up a chance for a particularly insightful
quantitative assessment of the potential of the purely relational
approach to \XQuery{} processing discussed here: not only can a
comparison with \pureXML{} be performed on the same machine, but even in
the context of a single query processing infrastructure (implementation
of query optimizer, plan operators, and B-tree indexes).

\begin{table}[t]
  \centering\small
  \renewcommand{\arraystretch}{1.22}
  \caption{Sample query set 
    taken from \protect\cite{TurboXPath} (rightmost
    column shows the query identifier used 
    in \protect\cite{TurboXPath}). We replaced the non-standard 
    \xquery{return-tuple}~\eqref{Q6} by an
    \SQL/\XML{} \xquery{XMLTABLE} construct.}
  \label{turboxpath-queries}
  \vspace{-2ex}
  \begin{tabular}[t]{@{}cll@{~}c@{}}
    \toprule         
    \multicolumn{2}{c}{\textbf{Query}} &
    \multicolumn{1}{c}{\textbf{Data}} &
    \textbf{\cite{TurboXPath}}
    \\
    \cmidrule(r){1-2}
    \cmidrule(lr){3-3}
    \cmidrule(l){4-4}
    \lbl{query}{$Q_3$}{Q3} &
    \begin{BVerbatim}[baseline=t]
/site/people/person[@id = "person0"]
            /name/text()
    \end{BVerbatim}
    &
    XMark & 9a \\[2ex]
    \lbl{query}{$Q_4$}{Q4} &
    \xquery{//closed\_auction/price/text()} &
    XMark & 9c \\[1ex]
    \lbl{query}{$Q_5$}{Q5} &
    \begin{BVerbatim}[baseline=t]
/dblp/*[@key = "conf/vldb2001" and
        editor and title]/title
    \end{BVerbatim}
    &
    DBLP & 8c \\[2ex]
    \lbl{query}{$Q_6$}{Q6} &
    \begin{BVerbatim}[baseline=t,commandchars=\\\{\},codes={\catcode`$=3\catcode`\_=8}]
for \vardollar{}thesis in /dblp/phdthesis
                [year < "1994" and
                 author and title]
return-tuple \vardollar{}thesis/title,
             \vardollar{}thesis/author,
             \vardollar{}thesis/year
    \end{BVerbatim}
    &
    DBLP & 8g 
    \\
    \bottomrule
  \end{tabular}
\end{table}

\begin{table}[t]
  \centering\small
  \caption{Observed result sizes and wall clock execution times 
    (average of 10~runs).}   
  \label{experimental-results}
  \vspace{-2ex}
  $
  \begin{array}[t]{@{}cr....@{}}
    \toprule 
    \multicolumn{2}{c}{\textbf{Query}} &
    \multicolumn{2}{c}{\textbf{DB2 + \Pathfinder{}}} &
    \multicolumn{2}{c}{\textbf{DB2\,\pureXML}}
    \\
    & & 
    \multicolumn{1}{c}{\text{stacked}} &
    \multicolumn{1}{c}{\textbf{join graph}} &
    \multicolumn{1}{c}{\text{whole}} & 
    \multicolumn{1}{c}{\text{segmented}} \\ 
    & \text{\#\,nodes} &
    \multicolumn{1}{c}{\StopWatchStart\,\text{(sec)}} & 
    \multicolumn{1}{c}{\StopWatchStart\,\text{(sec)}} & 
    \multicolumn{1}{c}{\StopWatchStart\,\text{(sec)}} &
    \multicolumn{1}{c}{\StopWatchStart\,\text{(sec)}} \\ 
    \cmidrule(r){1-2}     
    \cmidrule(lr){3-4}
    \cmidrule(l){5-6}
      \text{\ref{Q1}}  & 1,625,157 &  63.011 & \mathbf{11}.\mathbf{788} & 10.073 & 9.661 \\
      \text{\ref{Q2}}  & 318       &
      \multicolumn{1}{c}{\quad\textsc{dnf}}  & 
      \mathbf{0}.\mathbf{544} &  
      \multicolumn{1}{c}{\quad\textsc{dnf}}  & 
      \multicolumn{1}{c}{\quad~~\textsc{dnf}} \\[2ex]
      \text{\ref{Q3}}  & 1         &  60.582 &  \mathbf{0}.\mathbf{017} & 0.891  & 0.001 \\
      \text{\ref{Q4}}  &     9,750 &  32.246 &  \mathbf{0}.\mathbf{309} & 6.455  & 7.438 \\
      \text{\ref{Q5}}  & 1         & 442.745 &  \mathbf{0}.\mathbf{391} & 48.066 & 0.001 \\
      \text{\ref{Q6}}  & 59        &   0.026 &  \mathbf{0}.\mathbf{004} & 1.292  & 0.017 
    \\
    \bottomrule
  \end{array}
  $
\end{table}

\smallskip\noindent
In a \pureXML{}-enabled database, \XML{} documents are held in columns
of type \sql{XML}, containing references to pages which store a
fragmented tree-structured infoset representation.
Index-based \XQuery{}
processing is based on \sql{XMLPATTERN} indexes whose entries are the
typed values---cast to a \SQL{} data type like \sql{DOUBLE} or
\sql{VARCHAR($n$)}---of nodes that are selected by a given non-branching
\XPath{} location path of forward steps (along the \xquery{descendant},
\xquery{child}, and \xquery{attribute} axes). A new plan operator
primitive, $\col{XISCAN}$, can deploy such an index to evaluate a
general value comparison of an \XPath{} expression with a literal value
(provided that the indexed path covers the queried path \cite[see
\emph{index eligibility}]{pureXMLEligibility}). An index scan via
$\col{XISCAN}$ yields \emph{RIDs}, the identifiers of those rows that contain
\XML{} documents with matching nodes. Once these documents are retrieved,
their node trees need to be traversed to navigate to the matching nodes. This
actual tree traversal is implemented in terms of the new $\col{XSCAN}$
primitive.

Although a B-tree with key $\col{vnlkp}$ resembles an \sql{XMLPATTERN}
index of type \sql{VARCHAR($n$)}---both index (the string values of)
atomized \XML{} nodes---the B-tree is deployed quite differently: a
lookup yields document order ranks (column $\col{pre}$) which may
be used 
\begin{inparaenum}[(1)]
\item to \emph{directly access} the infoset properties of only the matching
  nodes via index $\col{p\ixinclude{}nvkls}$ or
\item to continue with \XPath{} location step processing.
\end{inparaenum}

To limit the overhead of the post-index-lookup tree traversal performed
by $\col{XSCAN}$, \pureXML{} favors database designs that lead to
comparably small \XML{} document segments (of a few KB, say) per row.
Since the tabular \XML{} infoset encoding and \XQuery{} processing
strategy discussed here can perfectly cope with very large \XML{}
instances (beyond 100\,MB), for the sake of comparison we let \pureXML{}
operate over both, segmented as well as monolithic \XML{} documents.

The internals of $\col{XSCAN}$ are based on the
Turbo\XPath{} algorithm \cite{TurboXPath}.  This is interesting in its 
own right: Turbo\XPath{} supports an \XPath~2.0-style dialect quite
similar to Fig.~\ref{grammar} and, in particular, admits nested 
\xquery{for} loops and \XPath{} predicates, but does not implement the
\emph{full axis} feature (Turbo\XPath{} supports the vertical axes).
Aware of $\col{XSCAN}$'s innards, to~\ref{Q1} and~\ref{Q2} 
we added sample queries that directly stem from \cite{TurboXPath}---these 
queries are displayed in Table~\ref{turboxpath-queries}.  
The resulting query set exhibited runtime characteristics representative
for the much larger query set we investigated in the course of this
work.

The Queries~\ref{Q1}--\ref{Q4} ran against an XMark instance of 110\,MB
(4,690,648~nodes), \ref{Q5}, \ref{Q6} queried an \XML{} representation
of Michael Ley's DBLP publication database (400\,MB or
31,788,688~nodes). We accommodated pure\XML{}'s preference for many but
smaller \XML{} documents and cut the whole XMark instance into 23,000
segments of 1--6\,KB; the DBLP instance was segmented into distinct
publications, yielding about 1,000,000 segments of 30~nodes (ca.\
800\,bytes) per row. To support the \pureXML{} processing
model, we further created an extensive \sql{XMLPATTERN} index family such that
at least one index was eligible to support the value references
occurring in the query set (\emph{e.g.}, for~$Q_3$ we created
an index on $\xquery{/site/categories/category/@id}$).

We then translated the query set with \Pathfinder{}, an \XQuery{}
compiler that includes a faithful implementation of loop lifting and
join graph isolation described in
Sections~\ref{loop-lifting-xquery-compiler}
and~\ref{xquery-join-graph-isolation}. \Pathfinder{} was configured to
emit the \SQL{} code derived from both, the original stacked plan and
the isolated join graph. The resulting \SQL{} queries ran against a
database populated with tabular \XML{} infoset encodings of the XMark
and DBLP instances, using a B-tree index setup as described in
Table~\ref{index-proposal}. Both, \pureXML{} and \Pathfinder{} used the
same DB2\,UDB\,V9.1 instance hosted on a dual 3.2\,GHz
Intel~Xeon\texttrademark{} computer with 8\,GB of primary and SCSI-based
secondary disk memory, running a Linux~2.6 kernel.

\smallskip\noindent
\textbf{The impact of join graph isolation.}
Table~\ref{experimental-results} summarizes the average wall clock
execution times we observed. For Query~\ref{Q1}, for example, isolating
the join graph (Fig.~\ref{plan-Q1-opt}) from the initial stacked plan
(Fig.~\ref{plan-Q1}) yields a five-fold reduction of execution time.
Compositional compilation leads to tall stacked plans that exhibit a
significant number of intermediate $\rank$ and $\distinct$ operators.
\Pathfinder{} translates such plans into a \SQL{} common table
expression (\sql{WITH\,$\dots$}) that features an equally large number
of \DISTINCT{} and \sql{RANK() OVER (\ORDERBY\,$\dots$)} clauses. Given
this, DB2\,V9 generates execution plans with numerous $\col{SORT}$
primitives followed by temporary table scans. This is different for the
join-graph-derived plans which compactly encode the \XQuery{} duplicate
and order semantics in terms of a single $\col{SORT}$ operator 
(see Figures~\ref{join-plan-Q1} and~\ref{join-plan-Q2}).
Similar and even more drastic effects could be observed for the other
queries: join graph isolation lets~\ref{Q2} execute in about 
$\nicefrac{1}{2}$~second (formerly the query did not complete within
20~hours), while the execution time of~\ref{Q3}--\ref{Q5} improved by
two to three orders of magnitude. Query~\ref{Q6} performs an early very
selective tag name test (\xquery{phdthesis}) and thus showed an
improvement of only a factor of six.

\smallskip\noindent
Table~\ref{experimental-results} further assesses how DB2\,\pureXML{}
fares against its \Pathfinder{}-driven relational self.
For~\ref{Q1}, the
universally high execution times of 
about 10~seconds largely reflect the substantial effort
to support serialization: the resulting node sequence contains
3,249 \xquery{open\_auction} elements, each being the root of a subtree 
containing 500~nodes on average.
Query~\ref{Q4} primarily relies on raw path traversal performance as no
value-based index can save the query engine from visiting a significant
part of the \XML{} instance. The more than 20-fold advantage of
\Pathfinder{} suggests that B-tree-supported location step
evaluation will remain a true challenger for the $\col{XSCAN}$-based
implementation inside \pureXML{}. Queries~\ref{Q3}, \ref{Q5}
(and~\ref{Q6}, to some extent) yield singleton (short) node sequences and
constitute the best case for the segmented \pureXML{} setup: here,
\sql{XMLPATTERN} index lookups return a single or few RID(s), directly
leading the system to small \XML{} segment(s)---the remaining traversal
effort for $\col{XSCAN}$ then is marginal. For the whole document
setup, however, an index lookup could only point to the single
monolithic \XML{} instance: $\col{XSCAN}$ thus does all the heavy work
(the wildcard \xquery{*} in~\ref{Q5} forces the engine to scan the
entire 400\,MB DBLP instance).  Despite the extensive index options
available to support~\ref{Q2}, \pureXML{} is not
able to finish evaluation within 20~hours: the system appears to miss
the opportunity to perform value-based selections and joins early 
(recall the discussion of Fig.~\ref{join-plan-Q2}) and ultimately is
overwhelmed by the Cartesian product of all \xquery{closed\_auction},
\xquery{item}, and \xquery{category} elements.  The indexes largely 
remain
unused (the predicate \xquery{price[data(.)\,>\,500]} is, in fact, 
evaluated second to last in the execution plan generated by \pureXML{}).

The sub-second execution times observed for \Pathfinder{} indicate
that the effort to compile into particularly simply-shaped self-join
chains pays off. The DB2\,V9 built-in monitor facility provides further
evidence in this respect: the queries enjoy a buffer cache hit ratio
of more than 90\,\% since merely table~$\doc$ and indexes fight for 
page slots.

\section{More Related Work}
\label{more-related-work}

One key ingredient in the join graph isolation process are the rewrites that
move order maintenance and duplicate elimination into tail positions.  Their
importance is underlined by similar optimizations proposed by other research
groups \cite{xqdup, timber, rainbow}: Fern{\'a}ndez et al. remove order
constraints and duplicates based on the \XQuery{} Core representation
\cite{xqdup}---an effect achieved by Rules~\eqref{rewrite-2} and
\eqref{rewrite-6} of Figure~\ref{rewrite-rules}.  The principal data structure
of \XQuery{} Core---item sequences---however prohibits merging of multiple
orders as in Rule~\eqref{rewrite-17}.

In \cite{rainbow}, a algebra working on ordered tables is the subject of order
optimization. An order context framework provides \emph{minimal ordered
semantics} by removing---much like Rule~\eqref{rewrite-2}---superfluous
\emph{Sortby} operators. In addition, order is merged in join operators and
pushed through the plan in an \emph{Orderby Pull up} much like in
Section~\ref{xquery-join-graph-isolation}.  In the presence of
\emph{order-destroying} operators such as $\distinct$ they however fail to
propagate the order information to the plan tail (compare to
Rule~\eqref{rewrite-14}).

An extension of the tree algebra (TLC-C) in the research project Timber 
introduces order on a global level \cite{timber} and generates tree
algebra plans that---if tree patterns are mapped to self-join chains---might
be transformed into \SQL{} queries similar to join graph isolation.

In Section~\ref{xpath-continuations}, we have seen how a
selectivity-based reordering of \XPath{} location steps can also lead to
a reversal of axes. In effect, the optimizer mimics a family of rewrites
that has been developed in \cite{LookingForward}. These rewrites were
originally designed to trade reverse \XPath{} axes for their forward
duals, which can significantly enlarge the class of expressions
tractable by streaming \XPath{} evaluators. Here, instead, we have found
the optimizer to exploit the duality in both directions---in fact, a
\xquery{descendant} axis step has been traded for an \xquery{ancestor}
step in the execution plan for~\ref{Q2} (Fig.~\ref{join-plan-Q2}). The
evaluation of rooted \xquery{/descendant::$n$} steps---pervasively
introduced in~\cite{LookingForward} to establish a context node set of
all elements with tag $n$ in a document---is readily supported by the
$\col{n}$-prefixed B-tree indexes. Since the \XQuery{} compiler
implements the \emph{full axis} feature, it can actually realize a
significant fraction of the rewrites in~\cite{LookingForward}.

Although we exclusively rely on the vanilla B-tree indexes that are
provided by any RDBMS kernel, cost-based join tree planning and join
reordering leads to a remarkable plan versatility. In the terminology of
\cite{QueryOptimizationforXML}, we have observed the optimizer to
generate the whole variety of \emph{Scan} (strict left-to-right location
path evaluation), \emph{Lindex} (right-to-left evaluation), and
\emph{Bindex} plans (hybrid evaluation, originating in a context node set
established via tag name selection; cf.\ the initial
\xquery{closed\_auction} node test in Fig.~\ref{join-plan-Q2}).

The path branching and stitching capability
(Section~\ref{xpath-continuations}) makes the present \XQuery{}
compilation technique a distant relative of the larger family of
\emph{holistic twig join} algorithms
\cite{HolisticTwigJoin,Twig2Stack,GTP}. We share the language dialect
of Fig.~\ref{grammar}---coined \emph{generalized tree pattern} queries
in~\cite{HolisticTwigJoin,GTP}---but add to this the \emph{full axis}
feature.
Quite differently, though, we 
\begin{inparaenum}[(1)]
\item let the RDBMS shoulder 100\,\% of the evaluation-time and parts of
  the compile-time effort invested by these algorithms (\emph{e.g.}, the
  join tree planner implements the $\emph{findOrder}(\cdot)$ procedure
  of~\cite{GTP} for free), and
\item use built-in B-tree indexes over table-shaped data where
  \emph{TwigStack}~\cite{HolisticTwigJoin} and
  \emph{Twig$^{\mathit{2}}${\kern-2pt}Stack}~\cite{Twig2Stack} rely on
  special-purpose runtime data
  structures, \emph{e.g.}, chains or hierarchies of linked stacks and
  modified B-trees, which call for significant invasive extensions to
  off-the-shelf database kernels.      
\end{inparaenum}

Finally, this work may be read as one possible response to a list of
open issues identified in the context of cost-based \XQuery{} processing
with DB2\,V9\,\pureXML{} (here, we directly refer to the specific issues 
raised in \cite[page~316\,ff.]{pureXMLCostBased}):
\setdefaultleftmargin{1em}{}{}{}{}{}
\begin{compactdesc}
\item[\normalfont\emph{\XML{} index exploitation}] 
  The infoset encoding (Fig.~\ref{encoded-XML}) is truly
  node-based: location paths may originate in any individual node 
  and the document order rank (column~$\col{pre}$) is sufficient to
  ``point into'' a document.
\item[\normalfont\emph{Deferred \XPath{} evaluation}] 
  The B-tree index keys (Table~\ref{index-proposal}) are
  self-contained: \XPath{} processing may resume without any
  consultation of a document's infoset encoding.  
\item[\normalfont\emph{Cost estimation}] 
  Since the generated executions plans exclusively feature
  the well-known operators (Table~\ref{lolepops}), established
  procedures for statistics collection and cost estimation
  remain applicable~\cite{CardForecast} (but see Section~\ref{work-in-flux}).
\item[\normalfont\emph{Order optimization}]  
  Join graph isolation leads to a sound---also in
  the presence of nested \xquery{for} loops---yet compact
  representation of the \XQuery{} order semantics in the plan tail,
  which ultimately finds its way into a single simple \SQL{} \ORDERBY{} clause. 
\end{compactdesc}
                 
\begin{figure*}[t]
\centering\small
\mprset{flushleft}
$$
\vspace{-1ex}
\begin{array}{@{}c@{}}
\inferrule
  {\relax}
  {
    \begin{array}{@{}c@{}}
      \env; \LOOP \vdash \xquery{doc(}\var{uri}\xquery) \Mapsto
    \\
      \pi_{\col{iter},\col{pos},\col{item}:\col{pre}} \left(
         \select_{(\col{kind}=\kind{DOC}) \wedge 
         (\col{name}=\var{uri})} (\doc)
         \cross 
         \attach_{\col{pos}:1}(\LOOP)
      \right)
    \end{array}
  }(\lbl{compile}{\textsc{Doc}}{compile-Doc})
\quad
\inferrule
  {\env; \LOOP \vdash e \Mapsto q}
  {
    \env; \LOOP \vdash \xquery{fs:ddo(}e\xquery) \Mapsto
    \rank_{\col{pos}:\langle\col{item}\rangle}\left(\distinct(
      \pi_{\col{iter},\col{item}}(q))\right)
  }(\lbl{compile}{\textsc{Ddo}}{compile-Ddo})
\\[10mm]
\inferrule
  {
    \begin{array}{@{}c@{}}
      \begin{array}{@{}l@{\qquad}l@{}}
         \{\dots, \xquery{\$}v_i \mapsto q_{v_i}, \dots\}; \LOOP \vdash e_{\var{if}} \Mapsto q_{\var{if}}
      &
        \LOOP_{\var{if}} \equiv 
        \distinct\left(\pi_{\col{iter_1}:\col{iter}}
                                              (q_{\var{if}})\right)
      \end{array}
    \\
      \left\{\dots,
             \xquery{\$}v_i \mapsto
             \pi_{\col{iter},\col{pos},\col{item}}
             (\LOOP_{\var{if}}
                   \Join_{\col{iter_1} = \col{iter}}
                   q_{v_i}),
             \dots
      \right\};
      \LOOP_{\var{if}} \vdash e_{\var{then}} \Mapsto q
    \end{array}
  }
  {
   \{\dots, \xquery{\$}v_i \mapsto q_{v_i}, \dots\}; \LOOP \vdash
    \xquery{if\,(fn:boolean($e_{\var{if}}$))\,then\,$e_{\var{then}}$\,else\,()} 
   \Mapsto q
  }(\lbl{compile}{\textsc{If}}{compile-If-Then})
\quad
\inferrule
  {
    \varolessthan \in 
    \{ \xquery{=},\xquery{!=},\xquery{<},\xquery{<=},\xquery{>},\xquery{>=}\} 
    \and
    \env; \LOOP \vdash e \Mapsto q
  }
  {
    \begin{array}{@{}c@{}}
    \env; \LOOP \vdash e \varolessthan \var{val} \Mapsto 
    \\
    \mathclap{
    \attach_{\col{item}:1}\left(
      \attach_{\col{pos}:1}\left(
        \distinct(\pi_{\col{iter}}(
          \sigma_{\col{value} \varolessthan \var{val}}(\doc \Join_{\col{pre} = \col{item}} q)
        ))
      \right)
    \right)}
    \end{array}
  }(\lbl{compile}{\textsc{Comp}}{compile-Comp})
\\[7mm]
\inferrule
  {
    \begin{array}{@{}c@{}}
      \{\dots, \xquery{\$}v_i \mapsto q_{v_i}, \dots\}; \LOOP \vdash e_{\var{in}} \Mapsto q_{\var{in}}
    \\
      q_{\var{\xquery{\$}x}} \equiv \num_{\col{inner}}(q_{\var{in}})
      \qquad
      \map \equiv \pi_{\col{outer}:\col{iter},
                         \col{inner},\col{sort}:\col{pos}}~(q_{\var{\xquery{\$}x}})
    \\
      \begin{array}{@{}l@{}}
      \left\{\dots,
             \xquery{\$}v_i \mapsto
             \pi_{\col{iter}:\col{inner},\col{pos},\col{item}}
               \left(\map \Join_{\col{outer} = \col{iter}} q_{v_i}\right),
             \dots
      \right\}
      +
    \\
      \left\{\xquery{\$}x \mapsto 
             \attach_{\col{pos}:1}
             \left(
               \pi_{\col{iter}:\col{inner},\col{item}}(q_{\var{\xquery{\$}x}})
             \right)
      \right\};
      \pi_{\col{iter}:\col{inner}}(\map)
      \vdash e_{\var{ret}} \Mapsto q
    \end{array}
    \end{array}
  }
  {
    \begin{array}{@{}c@{}}
       \{\dots, \xquery{\$}v_i \mapsto q_{v_i}, \dots\}; \LOOP \vdash
       \xquery{for\,\$$x$\,in\,$e_{\var{in}}$\,return\,$e_{\var{ret}}$}
       \Mapsto
     \\
       \pi_{\col{iter}:\col{outer},\col{pos}:\col{pos_1},\col{item}}
         \left(\rank_{\col{pos_1}:\langle\col{sort},\col{pos}\rangle}
                 \left(q
                       \Join_{\col{iter}=\col{inner}}
                       \map
                 \right)
         \right)
   \end{array}
  }(\lbl{compile}{\textsc{For}}{compile-For})
\quad
\inferrule
  {\relax}
  {\left\{\dots,\xquery{\$}x\mapsto q,\dots\right\}; \LOOP \vdash
   \xquery{\$}x\Mapsto q
  }(\lbl{compile}{\textsc{Var}}{compile-Var})
\\[10mm]
\inferrule
  { \env; \LOOP \vdash e \Mapsto q }
  {
    \begin{array}{@{}c@{}}
      \env; \LOOP \vdash e\xquery{/}\alpha\xquery{::}n \Mapsto
    \\
      \rank_{\col{pos}:\langle\col{item}\rangle} 
        \left(\pi_{\col{iter},\col{item}:\col{pre}}
              \left(\select_{\var{kindt}(n) \wedge
                             \var{namet}(n)}(\doc)
              \Join_{\var{axis}(\alpha)}
              \left(
                \pi_{\col{iter},
                     \col{pre}_{\circ}:\col{pre},
                     \col{size}_{\circ}:\col{size},
                     \col{level}_{\circ}:\col{level}}
                  \left( \doc \Join_{\col{pre} = \col{item}} q\right)
              \right)
        \right)
        \right)
    \end{array}
  }(\lbl{compile}{\textsc{Step}}{compile-Step})
\end{array}
$$
\addtocounter{figure}{1}
\caption{Rules defining the compilation scheme 
  $\Gamma;\LOOP \vdash e \Mapsto q$ from \XQuery{} expression~$e$ to 
  algebraic plan~$q$.}
\label{rules}
\addtocounter{figure}{-2}
\end{figure*}

\section{Work in Flux}
\label{work-in-flux}

This work rests on the maturity and versatility of
data\-ba\-se technology for strictly table-shaped data, resulting from 30+
years of experience. We 
\begin{inparaenum}[(1)]
\item discussed relational encodings of
  the true \XQuery{} semantics that are accessible for today's \SQL{}
  query optimizers, but
\item also saw that some care is needed to unlock the
  potential of a set-oriented query processor.
\end{inparaenum}

In the context of the open-source \Pathfinder{} project, we 
continue to pursue the idea of a purely relational \XQuery{} processor.
On the workbench lie DB2's \emph{statistical views}---in our case, pre-formulated
\xquery{descendant} and \xquery{child} location steps for which the
system records statistical properties but not the result itself---which
promise to give insight into the \emph{structural} node distribution of
an encoded \XML{} document. This may further improve join tree planning.

The scope of this work reaches beyond \XQuery{}. Tall stacked plan
shapes with scattered distributions of $\rank$ operators
(Fig.~\ref{plan-Q1}) also are an artifact of the compilation of complex
\SQL/OLAP queries (in which functions of the \sql{RANK()} family are
pervasive). The observations of Section~\ref{db2-udb-xquery-wizard-builtin}
suggest that the rewriting procedure of Fig.~\ref{rewrite-rules} can
benefit commercial query optimizers also in this domain.

\medskip\noindent \textbf{Acknowledgments.} This research is supported
by the German Research Council (DFG) under grant GR\,2036/2-1.

\newpage


             
\bibliographystyle{abbrv}
\baselineskip9.5pt
\bibliography{join-graph}

\begin{thebibliography}{10}

\bibitem{DB2V9}
{DB2\,9} for {Linux}, {UNIX} and {Windows} {Manuals}, 2007.
\newblock \url{http://www.ibm.com/software/data/db2/udb/}.

\bibitem{pureXMLEligibility}
A.~Balmin, K.~Beyer, F.~{\"O}zcan, and M.~Nicola.
\newblock {{On} the {Path} to {Efficient} {XML} {Queries}}.
\newblock In {\em Proc.\ VLDB}, 2006.

\bibitem{pureXMLCostBased}
A.~Balmin, T.~Eliaz, J.~Hornbrook, L.~Lim, G.~Lohman, D.~Simmen, M.~Wang, and
  C.~Zhang.
\newblock {Cost-based} {Optimization} in {DB2} {XML}.
\newblock {\em IBM Systems Journal}, 45(2), 2006.

\bibitem{W3C:XQuery}
S.~Boag, D.~Chamberlin, and M.~Fern{\'a}ndez.
\newblock {XQuery} 1.0: {An} {XML} {Query} {Language}.
\newblock W3 Consortium, 2007.
\newblock \url{http://www.w3.org/TR/xquery/}.

\bibitem{HolisticTwigJoin}
N.~Bruno, N.~Koudas, and D.~Srivastava.
\newblock {Holistic} {Twig} {Joins}: {Optimal} {XML} {Pattern} {Matching}.
\newblock In {\em Proc.\ SIGMOD}, 2002.

\bibitem{Twig2Stack}
S.~Chen, H.~Li, J.~Tatemura, W.~Hsiung, D.~Agrawal, and K.~Sel{\c c}uk~Candan.
\newblock {Twig{$^2$}Stack}: {Bottom-Up} {Processing} of
  {Generalized-Tree-Pattern} {Queries} over {XML} {Documents}.
\newblock In {\em Proc.\ VLDB}, 2006.

\bibitem{GTP}
Z.~Chen, H.~Jagadish, L.~Lakshmanan, and S.~Paparizos.
\newblock {From} {Tree} {Patterns} to {Generalized} {Tree} {Patterns}: {On}
  {Efficient} {Evaluation} of {XQuery}.
\newblock In {\em Proc.\ VLDB}, 2003.

\bibitem{W3C:XQueryFS}
D.~Draper, P.~Fankhauser, M.~Fern{\'a}ndez, A.~Malhotra, K.~Rose, M.~Rys,
  J.~Sim{\'e}on, and P.~Wadler.
\newblock {XQuery} 1.0 and {XPath} 2.0 {Formal} {Semantics}.
\newblock W3 Consortium, 2007.
\newblock \url{http://www.w3.org/TR/xquery-semantics/}.

\bibitem{xqdup}
M.~Fern{\'a}ndez, J.~Hidders, P.~Michiels, J.~Sim{\'e}on, and R.~Vercammen.
\newblock {Optimizing} {Sorting} and {Duplicate} {Elimination} in {XQuery}
  {Path} {Expressions}.
\newblock In {\em Proc.\ DEXA}, 2005.

\bibitem{PartitionedBtree}
G.~Graefe.
\newblock {Sorting} and {Indexing} with {Partitioned} {B-Trees}.
\newblock In {\em Proc.\ CIDR}, 2003.

\bibitem{XQueryOnSQLHosts}
T.~Grust, S.~Sakr, and J.~Teubner.
\newblock {XQuery} on {SQL} {Hosts}.
\newblock In {\em Proc.\ VLDB}, 2004.

\bibitem{XPathAccel}
T.~Grust, J.~Teubner, and M.~van Keulen.
\newblock {Accelerating} {XPath} {Evaluation} in {Any} {RDBMS}.
\newblock {\em ACM TODS}, 29(1), 2004.

\bibitem{TurboXPath}
V.~Josifovski, M.~Fontoura, and A.~Barta.
\newblock {Querying} {XML} {Streams}.
\newblock {\em VLDB Journal}, 14(2), 2004.

\bibitem{QueryOptimizationforXML}
J.~McHugh and J.~Widom.
\newblock {Query} {Optimization} for {XML}.
\newblock {\em VLDB Journal}, 1999.

\bibitem{TPoX}
M.~Nicola, I.~Kogan, and B.~Schiefer.
\newblock {An} {XML} {Transaction} {Processing} {Benchmark}.
\newblock In {\em Proc.\ SIGMOD}, 2007.

\bibitem{LookingForward}
D.~Olteanu, H.~Meuss, T.~Furche, and F.~Bry.
\newblock {XPath}: {Looking} {Forward}.
\newblock In {\em Proc.\ XMLDM (EDBT Workshop)}, 2002.

\bibitem{ORDPATH}
P.~E. O'Neil, E.~J. O'Neil, S.~Pal, I.~Cseri, G.~Schaller, and N.~Westbury.
\newblock {ORDPATHs}: {Insert-Friendly} {XML} {Node} {Labels}.
\newblock In {\em Proc.\ SIGMOD}, 2004.

\bibitem{timber}
S.~Paparizos and H.~V. Jagadish.
\newblock Pattern tree algebras: sets or sequences?
\newblock In {\em Proc.\ VLDB}, 2005.

\bibitem{XMark}
A.~Schmidt, F.~Waas, M.~Kersten, M.~Carey, I.~Manolescu, and R.~Busse.
\newblock {XMark:} {A} {Benchmark} for {XML} {Data} {Management}.
\newblock In {\em Proc.\ VLDB}, 2002.

\bibitem{CardForecast}
J.~Teubner, T.~Grust, S.~Maneth, and S.~Sakr.
\newblock {Dependable} {Cardinality} {Forecasts} for {XQuery}.
\newblock In {\em Proc.\ VLDB}, 2008.

\bibitem{rainbow}
S.~Wang, E.~A. Rundensteiner, and M.~Mani.
\newblock {Optimization} of {Nested} {XQuery} {Expressions} with {Orderby}
  {Clauses}.
\newblock {\em DKE Journal}.

\end{thebibliography}

\newpage 

\begin{appendix}

\subsection{Inference Rules}
\label{inference-rules}

The inference rule set of Fig.~\ref{rules} (adopted
from~\cite{XQueryOnSQLHosts}) implements a loop-lifting \XQuery{}
compiler for the \XQuery{} subset in Fig.~\ref{grammar} taking into
account the \XML{} encoding sketched in
Section~\ref{join-based-xquery-semantics}. The rule set defines a
judgment
$$
  \env;\LOOP \vdash e \Mapsto q \enskip,
$$ indicating
that the \XQuery{} expression $e$ compiles into the algebraic plan $q$,
given
\begin{compactenum}[(1)]
\item $\env$, an environment that maps \XQuery{} variables to their
  algebraic plan equivalent, and
\item $\LOOP$, a table with a single column $\col{iter}$ that invariantly
  contains $n$ arbitrary but distinct values if $e$ is evaluated in $n$
  loop iterations.
\end{compactenum}
An evaluation of the judgment
$\varnothing;
\raisebox{-0.75ex}{\smash[t]{$\begin{littbl}
  \begin{array}[b]{|c|}
    \colhd{\scriptstyle iter} \\
    \scriptstyle 1 \\
    \hline
  \end{array}
\end{littbl}$}} \vdash e_{0} \Mapsto q_{0}$
invokes the compiler for the top-level expression $e_{0}$ (the
singleton $\LOOP$ relation represents the single iteration
of a pseudo loop wrapped around $e_{0}$).  The inference
rules pass $\env$ and $\LOOP$ top-down and synthesize the plan $q_{0}$
in a bottom-up fashion.  A serialization operator at the plan root
completes the plan to read $\ser(q_{0})$. 

\end{appendix}
\end{document}